\documentclass[twocolumn,showpacs,preprintnumbers,amsmath,amssymb,
superscriptaddress]
{revtex4}
\usepackage{graphics}
\usepackage{epsfig}
\usepackage{dcolumn}
\usepackage{bm}
\newcommand{\be}{\begin{eqnarray}}
\newcommand{\ee}{\end{eqnarray}}

\renewcommand{\d}{\mbox{${\rm d}$}}

\begin{document}
\title{Second order brane cosmology with radion stabilization}
\author{G.L.~Alberghi}
\email{alberghi@bo.infn.it}
\affiliation{Department of Physics, University of Bologna,
Via Irnerio~46, 40126~Bologna, Italy.}
\affiliation{Department of Astronomy, University of Bologna, Italy.}
\affiliation{I.N.F.N., Sezione di Bologna, Italy.}
\author{D.~Bombardelli}
\email{bombardelli@bo.infn.it}
\affiliation{Department of Physics, University of Bologna,
Via Irnerio~46, 40126~Bologna, Italy.}
\affiliation{I.N.F.N., Sezione di Bologna, Italy.}
\author{R.~Casadio}
\email{casadio@bo.infn.it}
\affiliation{Department of Physics, University of Bologna,
Via Irnerio~46, 40126~Bologna, Italy.}
\affiliation{I.N.F.N., Sezione di Bologna, Italy.}
\author{A.~Tronconi}
\email{tronconi@bo.infn.it}
\affiliation{Department of Physics, University of Bologna,
Via Irnerio~46, 40126~Bologna, Italy.}
\affiliation{I.N.F.N., Sezione di Bologna, Italy.}
\begin{abstract}
We study cosmology in the five-dimensional Randall-Sundrum
brane-world with a stabilizing effective potential
for the radion and matter localized on the branes.
The analysis is performed by employing a perturbative
expansion in the ratio $\rho/V$ between the matter energy
density on the branes and the brane tensions around the
static Randall-Sundrum solution (which has $\rho=0$ and
brane tensions $\pm V$).
This approach ensures that the matter evolves adiabatically
and allows us to find approximate solutions to second
order in $\rho/V$.
Some particular cases are then analyzed in details.
\end{abstract}
\pacs{12.60.-i, 04.50.+h, 98.80.Cq, 98.80.-k}
\maketitle
\section{Introduction}
\label{I}
Higher and higher precision data which are about to be collected in
new experiments of particle physics and astrophysics in the next few
years convey considerable attention to theories with extra dimensions.
The main role of such theories, originally introduced in the 20's
by Kaluza and Klein~\cite{Kaluza:1921tu,Klein:1926tv}, is
to provide a connection between particle physics and gravity at some
level.
At a deeper level, string theory unifies all the interactions by
means of some $n$-dimensional manifold (with $n>4$) where the
fundamental objects are supposedly living; at a more phenomenological
level, models which assume the existence of extra dimensions,
no matter their origin, are considered in order to solve some
puzzles of particle physics, cosmology and astrophysics, giving
rise to many possible observable consequences.
\par
Originally proposed in order to solve the problem of the large
hierarchy between Gravity and Standard Model scales, the
Randall-Sundrum model of Ref.~\cite{Randall:1999ee} (RS~I)
has acquired considerable relevance due to its stringy
inspiration.
It represents the prototype of the so-called brane-world
and differs from previous models in that it constrains
standard matter on a four-dimensional manifold (the brane)
just letting gravity (and exotic matter) propagate everywhere.
The RS~I solution to the hierarchy problem needs one additional
compactified (orbifolded) spatial dimension with two branes
located at its fixed points, plus a negative cosmological
constant filling the space between such branes (the bulk).
The bulk cosmological constant $\Lambda$ warps the extra
dimension and generates the effective four-dimensional
physical constants we measure.
It was soon realized that the modifications to four-dimensional
gravity induced by the fifth dimension may be reduced to such
a short distance effect to be unobservable even in the presence
of just one brane and infinite compactification radius
(the RS~II model of Ref.~\cite{Randall:1999vf}).
\par
The cosmological features of the RS models are nowadays being
investigated even more than its particle physics consequences,
due to the refined results lately obtained and to the major
problems recent astrophysical data have revealed: the possible
late time acceleration from supernovae, CMBR spectrum, dark matter
and dark energy quests suggest either a full revision of the modern
theoretical physics approach or the possibility of the existence of
further, up to now ignored, ingredients such as the extra
dimensions.
\par
In particular the single brane RS~II cosmological
dynamics~\cite{Binetruy:1999ut,Binetruy:1999hy} is known to generate
$(\rho/V)^2$ corrections to standard Friedmann and acceleration
equations, where $\rho$ is the energy density of the fluid filling the
brane and $V$ is the constant brane tension.
These corrections are negligible when $\rho\ll V$,
the regime in which the RS~II model is reliable and
leads to standard cosmic evolution.
The two brane RS~I setup is much more involved: a stabilization
mechanism for the distance between the branes, such as that of
Ref.~\cite{Goldberger:1999uk}, is necessary to get the correct
hierarchy in the absence of matter.
Moreover, a bulk potential for the radion (the metric degree
of freedom associated with the fifth dimension) is necessary to
achieve solvable junction conditions when matter is present on the
boundaries~\cite{Csaki:1999mp}.
In this case, cosmological solutions to order
$\rho/V$~\cite{Cline:2000tx} are not sufficient to grasp
the particular features of the background metric evolution
originated by the extra dimension and one needs to investigate
the effect of terms of order $(\rho/V)^2$ (as was done
in Ref.~\cite{vinet}) or higher.
\par
The aim of this article is to go beyond the first order
approximation in brane cosmology for RS~I models with two
branes. 
Our approach will differ from Ref.~\cite{vinet}
in that we do not consider a bulk scalar field to stabilize
the radion but include an effective stabilizing potential
directly into the equations (see also Ref.~\cite{setare}).
Consequently, our perturbative expansion is around the
RS~I solution.
The calculations are then carried out in order to show how
$\rho^2$ contributions to the four-dimensional Hubble parameter
may affect the model (or may be unobservable).
Such terms are expected as fingerprints of the fifth dimension
in analogy with the single brane RS~II framework.
The latter case will also be studied as RS~I in the
limit when the distance between the branes
diverges.
Some hints about the possibility of an accelerated
expansion driven by exotic fluids with pressure $p=w\,\rho$
and $w>0$ will be presented, thus suggesting the necessity
to go beyond the second order approximation.
\par
The paper is organized as follows: in Section~\ref{II},
we present the complete setup of the model
under consideration;
in Section~\ref{III} the second order ansatz is described
and Einstein equations are perturbatively solved;
in Section~\ref{IV} cosmological consequences of the solutions
are analyzed and compared to the known brane-world solutions;
in Section~\ref{V} the analysis of the approximations is
performed and, finally, in Section~\ref{VI}, some conclusions
are drawn.
For the five-dimensional metric $g_{AB}$ we shall use
the signature $(+,-,-,-,-)$, so that
$g\equiv {\rm det}(g_{AB})>0$. 
\section{Einstein equations}
\label{II}
Let us consider a RS~I model perturbed by the 
presence of matter on the two branes.
The bulk metric is given by
\be
\d s^2&\equiv& g_{AB}\,\d x^A\,\d x^B
\nonumber
\\
&=&
n^2(y,t)\,\d t^2
-a^2(y,t)\,\d x^i\,\d x^i
-b^2(y,t)\,\d y^2
\ .
\label{metric}
\ee
The Einstein tensor for this metric is 
\begin{subequations}
\be
G_{00}\!\!\!\!&=&\!\!\!\!
3\left\{\left(\frac{\dot a}{a}\right)^2 +\frac{\dot a\, \dot  
b}{a\,b}-\frac{n^2}{b^2}
\left[\frac{a''}{a}+\left(\frac{a'}{a}\right)^2- 
\frac{a'\,b'}{a\,b}\right]\right\}
\label{G00}
\\
G_{ii}\!\!\!\!&=&\!\!\!\!
\frac{a^2}{b^2}\left[\left(\frac{a'}{a}\right)^2
+2\frac{a'\,n'}{a\,n}
-\frac{b'\,n'}{b\,n}-2\frac{a'\,b'}{a\,b}
+2\frac{a''}{a}+\frac{n''}{n}\right]
\nonumber
\\
\!\!
&&
\!\!\!\!\!\!
-\frac{a^2}{n^2}\left[\left(\frac{\dot a}{a}\right)^2
-2\frac{\dot a\,\dot n}{a\,n}
+2\frac{\ddot a}{a}-\frac{\dot b}{b}
\left(\frac{\dot n}{n}-2\frac{\dot a}{a}\right)
+\frac{\ddot b}{b}\right]
\label{Gii}
\\
G_{04}\!\!\!\!&=&\!\!\!\!
3\left[\frac{\dot a \,n'}{a\,n}+\frac{a'\,\dot b}{a\,b}
-\frac{\dot a'}{a}\right]
\label{G04}
\\
G_{44}\!\!\!\!&=&\!\!\!\!
3\left\{\frac{a'}{a}\left(\frac{a'}{a}+\frac{n'}{n}\right)- 
\frac{b^2}{n^2}\left[\frac{\dot a}{a}
\left(\frac{\dot a}{a}-\frac{\dot n}{n}\right)+\frac{\ddot  
a}{a}\right]\right\}
\label{G44}
\ee
\end{subequations}
where a prime denotes a derivative with respect to $y$ and
a dot a derivative with respect to the universal time $t$.
The energy-momentum tensor in the bulk is that of an
anti-de~Sitter space with the addition of a term generated
by a field which serves the purpose of stabilizing the distance
between the two branes of the RS~I model, 
\be
T^A_{\ B}=\Lambda\,g^A_{\ B}+\tilde T^A_{\ B}
\ ,
\ee
where, as usual,
\be
\tilde T_{AB}=
-\frac{2}{\sqrt{g}}\,
\frac{\delta\mathcal{L}_{\rm stab}}{\delta g^{AB}}
\ ,
\ee
and ${\mathcal{L}}_{\rm stab}$ is the Lagrangian of the
stabilizing field (for a scalar field, see
e.g.~Ref.~\cite{Goldberger:1999uk,Cline:2000tx,vinet}).
\par
We shall consider the particular case in which the
stabilizing mechanism can be effectively described
by a harmonic potential for the radion, with an
effective Lagrangian of the form~\cite{Csaki:1999mp}
\be
\mathcal{L}_{\rm eff}= -\sqrt{g}\,\omega^2\,(b-b_0)^2
\equiv-\sqrt{g}\,U(b)
\ ,
\label{effLag}
\ee
when the metric is written as in Eq.~(\ref{metric}),
and the potential $U$ depends on the component
$g_{44}$.
The two 3-branes have opposite tensions and their  
contribution to the total energy-momentum tensor is
given by
\be
T^A_{i\,B}\!\!&=&\!\!
\frac{\delta(y-y_i)}{b}
\nonumber
\\
&&
\times
{\rm diag}\left(V_i+\rho_i,V_i-p_i,V_i-p_i,V_i-p_i,0\right)
\ ,
\ee
where $i=p$, $n$, and $y_p=0$ ($y_n=1/2$) is the position
of the positive (negative) tension brane.
The Einstein equations in the bulk,
\be
G_{AB}=k^2\, T_{AB}
\ ,
\label{fullE}
\ee
form a system of four differential equations for the three
independent functions $f_\alpha=(n,a,b)$.
On using the Bianchi identity $\nabla_A G^{A0}=0$,
it is then straightforward to show that the three equations
\be
\left\{
\begin{array}{l}
G_{00}=k^2\,T_{00}
\\
\\
G_{04}=0
\\
\\
G_{44}=k^2\,T_{44}
\end{array}
\right.
\label{Ein}
\ee
are independent and form a complete set.
This means that a solution to Eqs.~(\ref{Ein}) also solves the full set~(\ref{fullE}).
Moreover, since $G_{AB}$ is trivially conserved because of the Bianchi identities,
the tensor $T_{AB}$ is also automatically conserved regardless of the
fact that the effective potential in Eq.~(\ref{effLag}) does not appear covariant.
\par
For computational purposes, it is convenient to do some further manipulation.
The first of Eqs.~(\ref{Ein}) can be replaced
by~\cite{Binetruy:1999ut,Binetruy:1999hy}
\be
F'(y,t)
+\frac{1}{6}\,k^2\,\left(\frac{\partial}{\partial y}  
a^4\right)\,T^0_{\;0}=0
\ ,
\label{EinEqs1}
\ee
with
\be
F(y,t)=\frac{(a\, a')^2}{b^2}-\frac{(a\,\dot a)^2}{n^2}
\ ,
\ee
which, on intergating along the extra dimension,
can be written as
\be
&&
\frac{\dot a^2}{a^2}
-\frac{n^2\,a'^2}{b^2\,a^2}
-\frac{k^2}{6}n^2\,T^0_{\ 0}
+\frac{k^2\,n^2}{6\,a^4}\,
\int\,a^4\,\left(T^0_{\ 0}\right)'\,\d y
\nonumber
\\
&&
=\frac{n^2}{a^4}\,\tilde c(t)
\ ,
\label{F00}
\ee
where $\tilde c(t)$ is related to the boundary conditions
at the branes.
The conservation equation $\nabla_A T^{A4}=0$ yields
\be
b'=(b_0-b)\,\left(\frac{n'}{n}
+3\frac{a'}{a}+2\frac{b'}{b}\right)
\ ,
\label{beq}
\ee
which is identically satisfied by the solutions of the
system~(\ref{Ein}).
Instead of the three Eqs.~(\ref{Ein}),
we shall therefore solve the equivalent system
\be
\left\{
\begin{array}{l}
\frac{\dot a^2}{a^2}
-\frac{n^2 a'^2}{b^2 a^2}
-\frac{k^2}{6}n^2 T^0_{\ 0}
+\frac{k^2 n^2}{6\,a^4}
\int a^4 \left(T^0_{\ 0}\right)'\d y
=\frac{n^2}{a^4} \tilde c(t)
\\
\\
b'=(b_0-b)\,\left(\frac{n'}{n}
+3\frac{a'}{a}+2\frac{b'}{b}\right)
\\
\\
G_{44}=k^2\,T_{44}
\ .
\end{array}
\right.
\label{Ein1}
\ee
\par
Moreover, bulk solutions must satisfy the boundary
equations given by the junction conditions on the two
branes,
\be
\left\{
\begin{array}{l}
\displaystyle\lim_{y\rightarrow y_i^+}
\frac{a'}{a}=\left.
-\frac{k^2}{6}\left(V_i+\rho_i\right)\,b\right|_{y=y_i}
\\
\\
\displaystyle\lim_{y\rightarrow y_i^+}
\frac{n'}{n}=\left.
-\frac{k^2}{6}\left[V_i-\left(2+3\,w_i\right)\rho_i\right]\,
b
\right|_{y=y_i}
\ ,
\end{array}
\right.
\label{juncon}
\ee
where we assumed an equation of state for the
vacuum perturbations of the form $p_i=w_i\,\rho_i$.
When $\rho_i\rightarrow 0$ the RS~I solution is fully
recovered and one finds the usual warped static
metric with $\tilde c(t)=0$ and
\be
\begin{array}{l}
n_{RS}(y,t)=a_{RS}(y,t)=\exp\left(-m\,b_0\,|y|\right)
\\
\\
b_{RS}=b_0
\ ,
\end{array}
\label{0order}
\ee
which also require the well known fine-tuning
\be
V_p=-V_n=\frac{6m}{k^2}
\ ,
\quad\quad
\Lambda=-\frac{6 m^2}{k^2}
\ .
\label{finetune}
\ee
\section{The Low Density Expansion}
\label{III}
A perturbative approach can be adopted in brane cosmology to  
investigate solutions to the Einstein equations by taking
as a starting point the static RS~I metric with $\rho_i=0$
reviewed in the previous section.
In fact, in the low density regime
\be
\frac{\rho_i}{|V_i|}\ll 1
\ ,
\ee
one can express the corrections to the solution~(\ref{0order})
to all orders in $\rho_i/V_i$ by assuming that the metric
functions $f_\alpha=(n,a,b)$ can be written
as~\footnote{Note that the approximation which makes use of an
effective Lagrangian is only compatible with second order
calculations.}
\be
f_\alpha=f_{RS}+\delta f_\alpha
\ ,
\ee
with $\delta f_\alpha\sim\sum_{n_i,n_j\ge 1}
c_{n_i n_j}\,\rho_i^{n_i}\,\rho_j^{n_j}$.
\par
In order to keep track of the various orders in the above
expansion, it is useful to introduce an expansion parameter
$\epsilon$ by replacing $\rho_i\to\epsilon \,\rho_i$
(and setting $\epsilon=1$ at the end of the computation).
We make the following {\em ansatz\/} for the metric,
\begin{subequations}
\be
&&
\!\!\!\!\!
n(y,t)=\exp\left(-m\,b_0\,|y|\right)\,
\left[1+\delta f_n(y,t)\right]
\label{exactansatz_n}
\\
\nonumber
\\
&&
\!\!\!\!\!
a(y,t)=a_h(t)\,\exp\left(-m\,b_0\,|y|\right)\,
\left[1+\delta f_a(y,t)\right]
\label{exactansatz_a}
\\
\nonumber
\\
&&
\!\!\!\!\!
b(y,t)=b_0+\delta f_b(y,t)
\ ,
\label{exactansatz_b}
\ee
\end{subequations}
so that the homogeneous scale factor $a_h(t)$ is
factored out, and $b_0$ is the equilibrium point corresponding
to the RS~I model [see Eq.~(\ref{0order}) above].
The solutions to the equations~(\ref{Ein1}) can then be
completely expressed in terms of the functions $\delta f_\alpha$
and $H_h\equiv{\dot a_h}/{a_h}$,
which we expand to second order in $\epsilon$ as
\begin{subequations}
\be
\delta f_\alpha&\simeq&
\epsilon\,
\left[f_{\alpha,p}^{(1)}(y)\,\rho_p
+f_{\alpha,n}^{(1)}(y)\,\rho_n\right]
\label{genexp}
\\
&&
+\epsilon^2\,
\left[f_{\alpha,p}^{(2)}(y)\,\rho_p^2
+f_{\alpha,n}^{(2)}(y)\,\rho_n^2+f_ 
{\alpha,m}^{(2)}(y)\,\rho_p\,\rho_n\right]
\nonumber
\\
H_h^2
&\simeq&
\epsilon\,\left(h_{h,p}^{(1)}\,\rho_p
+h_{h,n}^{(1)}\,\rho_n\right)
\nonumber
\\
&&
+\epsilon^2\,
\left(h_{h,p}^{(2)}\,\rho_{p}^2
+h_{h,n}^{(2)}\,\rho_n^2
+h_{h,m}^{(2)}\,\rho_p\,\rho_n\right)
\ .
\label{Hhom}
\ee
We also expand $c\equiv \tilde c/a_h^4$ as
\be
c(t)
&\simeq&
\epsilon\,
\left(c_{p}^{(1)}\,\rho_p+c_{n}^{(1)}\,\rho_n\right)
\nonumber
\\
&&
+\epsilon^2\,
\left(c_{p}^{(2)}\,\rho_p^2+c_{n}^{(2)}\,\rho_n^2
+c_{m}^{(2)}\,\rho_p\,\rho_n\right)
\ .
\label{cexp}
\ee
\end{subequations}
Note that the time dependence is just carried by the functions
$\rho_i=\rho_i(t)$ and that, of all the coefficients appearing
above, only the $f_{\alpha,i}^{(n)}$'s in Eq.~(\ref{genexp})
depend on $y$, whereas the others are constant.
\par
In order to proceed with the perturbative expansion, one also
needs to expand $\dot \rho_i$.
From the conservation equation~\footnote{Note that this
conservation equation can be obtained by taking the limit
$y\to y_i$ in the equation $G_{04}=0$.}
\be
\dot\rho_i=
-3\,H(y_i,t)\,\left(1+w_i\right)\,\rho_i
\ ,
\label{continuityeq}
\ee
it immediately follows that the time evolution
of the matter densities is adiabatic, since
$H\equiv \dot a/a\sim\rho^{1/2}$ and, therefore,
\be
\frac{|\dot\rho_i|}{\rho_i^{5/4}}
\sim
\left(\frac{\rho_i}{|V_i|}\right)^{1/4}
\ll 1
\ .
\ee
If we now assume that, to second order in $\epsilon$,
\be
H^2(y,t)
\!\!&\simeq&\!\!
\epsilon
\left[h_{p}^{(1)}(y)\,\rho_p
+h_{n}^{(1)}(y)\,\rho_n\right]
\label{Hexpansion}
\\
&&
+\epsilon^2
\left[h_{p}^{(2)}(y)\,\rho_{p}^2
+h_{n}^{(2)}(y)\,\rho_n^2+h_{m}^{(2)}(y) 
\,\rho_p\,\rho_n\right]
\ ,
\nonumber
\ee
the coefficients of the above expansion can be related
to the corresponding ones in Eqs.~(\ref{Hhom}) and
(\ref{genexp}) for $\alpha=a$ by equating  
the two expressions for $H^2$ at $y=y_i$ up to second
order,
\be
H^2(y_i,t)=\left(H_h(t)
+\frac{\dot{\delta f_a}(y_i,t)}{1+\delta f_a(y_i,t)}\right)^2
\ .
\ee
With the help of Eq.~(\ref{continuityeq}), one finally obtains
\begin{subequations}
\be
h_i^{(1)}(y)&=&h_{h,i}^{(1)}
\\
\nonumber
\\
h_i^{(2)}(y)&=&
h_{h,i}^{(2)}-6(1+w_i)\,h_{h,i}^{(1)}\,f_{a,i}^{(1)}(y)
\\
\nonumber
\\
h_m^{(2)}(y)&=&
h_{h,m}^{(2)}-6(1+w_p)\,h_{h,n}^{(1)}\,f_{a,p}^{(1)}(y) 
\nonumber
\\
&&
-6(1+w_n)\,h_{h,p}^{(1)}\,f_{a,n}^{(1)}(y)
\ .
\label{Hcoeff}
\ee
\end{subequations}
\subsection{First order results}
In order to solve the bulk equations order by order,
one has to substitute the previous expansion in
the dynamical equations (\ref{Ein1}).
This will allow us to determine explicitly the coefficients
$ \delta f_{\alpha}$ once the boundary conditions are imposed.
Let us begin with first order equations.
\par
At order $\epsilon $, the constraint (\ref{beq}) reads
\be
{f_{b,p}^{(1)}}'\rho_p
+{f_{b,n}^{(1)}}'\rho_n 
-4\,m\,b_0
\left[
f_{b,p}^{(1)}\,\rho_p
+f_{b,n}^{(1)}\,\rho_n
\right]=0
\ .
\ee
If we allow $\rho_p$ and $\rho_n$ to be arbitrary functions
of the time, the above equation splits into two independent
equations for the coefficients of $\rho_i$, and one finds
\be
f_{b,i}^{(1)}(y)=b_i^{(1)}\,\exp\left(4\,m\,b_0\,y\right)
\ ,
\label{fbi1}
\ee
where $b_i^{(1)}$ are constant coefficients
to be determined.
\par
The functions $f_{a,i}^{(1)}(y)$'s can now be calculated by
solving Eq.~(\ref{F00}). 
Since the contribution of the stabilizing potential vanishes
at order $\epsilon$, the integral-differential
equation~(\ref{F00}) becomes the first order linear
differential equation
\begin{widetext}
\be
\sum_i{\rm e}^{-2\,m\,b_0\,y}\,
\rho_i\left(2\,m^2\,b_i^{(1)}\,
{\rm e}^{4\,m\,b_0\,y}-b_0\,c_i^{(1)}\,{\rm e}^{4\,m\,b_0\,y}
+b_0\,h_{h,i}^{(1)}\,{\rm e}^{2\,m\,b_0\,y}
+2\,m\,{f_{a,i}^{(1)}}'\right)=0
\ .
\ee
On solving for each coefficient of $\rho_i$ independently,
one obtains
\be
f_{a,i}^{(1)}&=&\frac{1}{8}\,{\rm e}^{2\,m\,b_0\,y}\,
\left[{\rm e}^{2\,m\,b_0\,y}\,
\left(\frac{c_i^{(1)}}{m^2}
-2\,\frac{b_{i}^{(1)}}{b_0}\right)
-2\,\frac{h_{h,i}^{(1)}}{m^2}\right]
+c_{a,i}^{(1)}
\ .
\label{fai1}
\ee
Finally, one can expand the equation $G_{44}=k^2\,T_{44}$
to first order,
\be
\sum_i
\rho_i\left[
{\rm e}^{2\,m\,b_0\,y}
\left(6b_0h_{h,i}^{(1)}+9b_0w_ih_{h,i}^{(1)} 
-9b_0c_{i}^{(1)}{\rm e}^{2\,m\,b_0\,y}
-6m^2b_i^{(1)}{\rm e}^{2\,m\,b_0\,y}
+4k^2\omega^2b_0^2b_i^{(1)}{\rm e}^{2\,m\,b_0\,y}\right)
-6mf_{n,i}^{(1)}\right]
=0
\ ,
\ee
and solve the two equations for $f_{n,i}^{(1)}(y)$.
The result is 
\be
f_{n,i}^{(1)}=\frac{1}{2}\,{\rm e}^{2\,m\,b_0\,y}\,
\left[{\rm e}^{2\,m\,b_0\,y}\,
\left(\frac{k^2\,\omega^2}{3\,m^2}\,b_0\,b_i^{(1)}
-\frac{3\,c_i^{(1)}}{4\,m^2}- 
\frac{b_i^{(1)}}{2\,b_0}
+\frac{(2+3\,w_i)}{2\,m^2}\,h_{h,i}^{(1)}\right) 
\right]+c_{n,i}^{(1)}
\ .
\label{fni1}
\ee
\end{widetext}
\par
We are now left with six numerical coefficients
\begin{subequations}
\be
h_{h,i}^{(1)}
\ ,
\quad
b_i^{(1)}
\ ,
\quad
c_i^{(1)}
\ ,
\ee
and four integration constants
\be
c_{a,i}^{(1)}
\ ,
\quad
c_{n,i}^{(1)}
\ .
\label{int_c}
\ee
\end{subequations}
In order to fix the above, one has to use the junction
conditions.
These four conditions, written in terms of the coefficients
of $\rho_i$, form a system of eight equations:
the discontinuity constraints for $a'(y,t)$ at $y=y_i$
imply
\be
\left\{
\begin{array}{l}
\rho_p\left(3\,c_p^{(1)} 
-3\,h_{h,p}^{(1)}+k^2\,m\right)+3\,\rho_n\left(c_n^{(1)}- 
h_{h,n}^{(1)}\right)=0
\\
\\
3\,\rho_p\left(c_p^{(1)}{\rm e}^{m\,b_0}-h_{h,p}^{(1)}\right)
\\
+\rho_n\left(3\,c_n^{(1)}{\rm e}^{2\,m\,b_0}
-3\,h_{h,n}^{(1)}{\rm e}^{m\,b_0}-k^2\,m\right)=0
\ ,
\end{array}
\right.
\ee
whose solution is given by
\begin{subequations}
\be
h_{h,p}^{(1)}=
\frac{k^2\,m\,{\rm e}^{m\,b_0}}{3\left({\rm e}^{m\,b_0}-1\right)}
\ ,
\quad
\quad
h_{h,n}^{(1)}=
\frac{k^2\,m\,{\rm e}^{-m\,b_0}}{3\left({\rm e}^{m\,b_0}-1\right)}
&&
\\
\nonumber
\\
c_p^{(1)}=\frac{k^2\,m}{3\left({\rm e}^{m\,b_0}-1\right)}
\ ,
\quad
\quad  
c_n^{(1)}=
\frac{k^2\,m\,{\rm e}^{-m\,b_0}}{3\left({\rm e}^{m\,b_0}-1\right)}
\ .
&&
\ee
\end{subequations}
The two analogous constraints for $n'(y,t)$ are both
equivalent to the equation
\be
&&
{\rm e}^{m\,b_0}\left[m\left(3\,w_p-1\right)+
4\,\omega^2\,b_0\,b_p^{(1)}
\left({\rm e}^{m\,b_0}-1\right)\right]\rho_p
\\
&&
+\left[m\left(3\,w_n-1\right)+
4\,\omega^2\,b_0\,b_n^{(1)}{\rm e}^{m\,b_0}
\left({\rm e}^{m\,b_0}-1\right)\right]\rho_n=0
\ ,
\nonumber
\ee
which yields
\be
&&
b_p^{(1)}=
\frac{m\,(3\,w_p-1)}{4\,\omega^2\,b_0\left({\rm e}^{m\,b_0}-1\right)}
\nonumber
\\
\\
&&
b_n^{(1)}=
\frac{m\,(3\,w_n-1)\,{\rm e}^{-m\,b_0}}{4\,\omega^2\,b_0
\left({\rm e}^{m\,b_0}-1\right)}
\nonumber
\ .
\ee
We see that the junction conditions are not sufficient
to determine the integration constants (\ref{int_c}).
Such freedom is in fact related to the gauge freedom in
the choice of the initial value for the scale factor and
time variable.
Without loss of generality, and to simplify the second order 
calculations, we then set $c_{a,i}^{(1)}=0$.
The values of the $c_{n,i}^{(1)}$'s are related to the choice of
the time variable.
Since one usually considers the negative tension brane in RS~I
as the four-dimensional ``visible Universe'',
it is natural to use the proper time $\tau$ on this brane
and choose the $c_{n,i}^{(1)}$'s so as to have
$n(y_n,\tau)=1$.
This can be achieved by setting
\begin{widetext}
\be
&&
c_{n,p}^{(1)}=
-\frac{{\rm e}^{2\,m\,b_0}\left[3\,m^2\,(3w_p-1)
+2\,b_0^2\,k^2\,\omega^2\,(3w_p+2)\right]}
{48\,m\,b_0^2\,\omega^2\left({\rm e}^{m\,b_0}-1\right)}
\nonumber
\\
\label{cn1negtime}
\\
&&
c_{n,n}^{(1)}=
\frac{{\rm  e}^{m\,b_0}\,
\left[3\,m^2\,(3w_n-1)+2\,b_0^2\,k^2\,\omega^2\,
\left(1-2{\rm  e}^{-m\,b_0}\right)\,
(3w_n+2)\right]}
{48\,m\,b_0^2\,\omega^2\,\left({\rm e}^{m\,b_0}-1\right)}
\ .
\nonumber
\ee
\end{widetext}
and defining the time coordinate $\tau$ as
\be
\d\tau=\exp\left(-\frac{m\,b_0}{2}\right)\,\d t
\ .
\ee
With this choice, the cosmological Friedmann equations can
be easily compared to standard ones.
\par
Let us now comment on the first order results. 
As far as the radion perturbation is concerned,
we found
\be
\delta f_b\!\!&=&\!\!
\frac{m\,{\rm e}^{4\,m\,b_0y}}{4\,b_0\,\omega^2}\,
\left[(1-3w_p)\,\rho_p
+{\rm e}^{-m\,b_0}(1-3w_n)\,\rho_n\right]
\epsilon
\nonumber
\\
&&
+\mathcal{O}(\epsilon^2)
\ ,
\label{dfb1}
\ee
which was expected, as it is due to the known coupling of
the radion with the trace of the energy-momentum tensor
of brane matter. 
Traceless fluids, such as radiation, have no first order
effect on the excitation of the radion. 
If one fills the branes with some pressureless fluid,
the distance between the two branes grows.
This effect, being counter-intuitive for the attractive
nature of Newtonian gravity, is in fact a consequence of
the form of the stabilizing potential.
Its non trivial contribution to the bulk energy-momentum
tensor at order $\epsilon$ is
\be
T_{44}\sim-2\,b_0^3\,\omega^2\,\delta f_b
\ .
\label{firstorderT}
\ee
The first order expressions are identical to those for a
static solution, as they can be obtained by neglecting
$\dot\rho_i$.
Every kind of matter on the branes thus acts so as to
detune the brane tensions from the bulk cosmological
constant and can be balanced by some constant pressure
along the fifth dimension.
Such pressure is given by the first order contribution
of (\ref{firstorderT}) which increases when $\delta f_b$
decreases.
\par
Note that (\ref{dfb1}) is proportional to the inverse
of $\omega^2$, which represents the effective spring constant
coming from some stabilization mechanism.
When such a constant diverges, the correction $\delta f _b$
vanishes and the length of the fifth dimension is fixed as
expected, even if there is a finite, $\omega$-independent  
$T_{44}$ pressure term.
On the other hand, the correction to the scale factor,
\begin{widetext}
\be
\delta f_a&=&
\frac{{\rm e}^{2\,m\,b_0\,y}}
{48\,m\,b_0^2\,\omega^2\,\left({\rm e}^{m\,b_0}-1\right)}
\left\{\left[2\,b_0^2\,k^2\,\omega^2\left({\rm e}^{2\,m\,b_0\,y}
-2\,{\rm e}^{m\,b_0}\right)
+3\,m^2\,(3w_p-1)\,{\rm e}^{2\,m\,b_0\,y}\right]\rho_p
\right.
\nonumber\\
&&
\left.
+{\rm e}^{-m\,b_0}\,
\left[2\,b_0^2\,k^2\,\omega^2
\left({\rm e}^{2\,m\,b_0\,y}-2\right)
+3\,m^2(3w_n-1){\rm e}^{2\,m\,b_0\,y}\right]
\rho_n
\right\}
\epsilon
+\mathcal{O}(\epsilon^2)
\ ,
\label{df1}
\ee
never vanishes when matter is present on the branes,
even if that is trace-less.
Furthermore a finite, non vanishing $\delta f_a$ can be obtained  
in the limit of infinite spring constant, regardless of the matter
equation of state.
\par
The carrection to the lapse function,
\be
\delta f_n\!\!&=&\!\!
\frac{{\rm e}^{2\,m\,b_0\,y}}
{48\,m\,b_0^2\,\omega^2\left({\rm e}^{m\,b_0}-1\right)}
\left\{
\left[2\,b_0^2\,k^2\,\omega^2\,(3w_p+2)
\left(2{\rm e}^{m\,b_0}-{\rm e}^{2\,m\,b_0\,y}\right)
+3\,m^2\,(3w_p-1)\,{\rm e}^{2\,m\,b_0\,y}
\right]\rho_p
\right.
\label{dfn1}
\\
&&
\left.
+{\rm e}^{-m\,b_0}\,
\left[2\,b_0^2\,k^2\,\omega^2\,(3w_n+2)\,
\left(2-{\rm e}^{2\,m\,b_0\,y}\right)
+3\,m^2(3w_n-1)\,{\rm e}^{2\,m\,b_0\,y}\right]
\rho_n
\right\}
\epsilon
+\epsilon\,c_{n,p}^{(1)}\,\rho_p
+\epsilon\,c_{n,n}^{(1)}\,\rho_n
+\mathcal{O}(\epsilon^2)
\ ,
\nonumber
\ee
\end{widetext}
does not vanish when the branes are filled with trace-less
matter.
Note, however, that a vanishing correction can be obtained
for some exotic fluid with $w_i=-2/3$ and negative pressure
in the limit $\omega\to\infty$.
Apart from these exceptions, one has non negligible
corrections everywhere in the bulk.
\par
In order to compare the first order results with the RS~II case
of a single brane, we must instead use the proper time on the
positive tension brane.
This is achieved by setting
\be
\!
c_{n,p}^{(1)}\!\!\!&=&\!\!\!
\frac{\left[3\,m^2(3w_p-1)
+2\,b_0^2\,k^2\,\omega^2
\left(1-2{\rm e}^{m\,b_0}\right)
(3w_p+2)\right]}
{48\,m\,b_0^2\,\omega^2\left({\rm e}^{m\,b_0}-1\right)}
\nonumber
\\
\label{cn1postime}
\\
\!
c_{n,n}^{(1)}\!\!\!&=&\!\!\!
\frac{{\rm e}^{-m\,b_0}\,
\left[3\,m^2(3w_n -1)
+2\,b_0^2\,k^2\,\omega^2\,(3w_n+2)\right]}
{48\,m\,b_0^2\,\omega^2\left(1-{\rm e}^{m\,b_0}\right)}
\ ,
\nonumber
\ee
and letting $b_0\to\infty$. 
The Friedmann equation is simply obtained by keeping
$\mathcal{O}(\epsilon)$ terms in $H^2(y_i,t)$. 
Since the first order four-dimensional Hubble parameter
is homogeneous, it reacts to all the sources along the
$y$ direction. 
On the positive tension brane one has, to first order
in $\epsilon$,
\be
H^2_p=\frac{m\,k^2\left({\rm e}^{m\,b_0}\rho_p
+{\rm e}^{-m\,b_0}\rho_n\right)}
{3\left({\rm e}^{m\,b_0}-1\right)}\,
\epsilon
\stackrel{b_0\to\infty}
{\longrightarrow}
\frac{m\,k^2}{3}\,\rho_p\,\epsilon
\ ,
\ee
regardless of the value of $\omega$. 
When the negative tension brane is moved to infinity,
its contribution goes to zero and one recovers the usual
first order effect in brane cosmology.
\par
On the visible brane, the Friedmann equation is slightly
modified by the rescaled time parameter $\tau$,
\be
H^2_n=\frac{m\,k^2\left({\rm e}^{2\,m\,b_0}\rho_p
+\rho_n\right)}
{3\left({\rm e}^{m\,b_0}-1\right)}\,\epsilon
\stackrel{\rho_p\to 0}{\longrightarrow}
\frac{m\,k^2\,\rho_n\,\epsilon}{3\left({\rm e}^{m\,b_0}-1\right)}
\ .
\ee
The acceleration equation is  
homogeneous as the Friedmann equation and becomes
\be
\frac{\ddot a(y_i,t)}{a(y_i,t)}&\!\!=\!\!&
\frac{m\,k^2\left[{\rm e}^{m\,b_0}(1+w_p)\rho_p
+{\rm e}^{-m\,b_0}(1+w_n)\rho_n\right]}
{6\left(1-{\rm e}^{m\,b_0}\right)}\epsilon
\ ,
\nonumber
\\
&&
\ee
which has the weighted brane fluid energy densities
as sources.
\subsection{Second order results}
We are now ready to evaluate $\mathcal{O}(\epsilon^2)$
corrections to the vacuum solution RS~I. 
The procedure will be analogous to the one used
for first order results in the previous section. 
\par
As a first step, one can impose the constraint (\ref{beq})
in order to find the dependence on $y$ of the second order
coefficients in $\delta f_b$.
We are then left with three inhomogeneous equations
obtained by setting to zero the coefficients of the
independent matter densities in
\begin{widetext}
\be
&&
{f_{b,p}^{(2)}}'\,\rho_p^2
+{f_{b,n}^{(2)}}'\,\rho_n^2
+{f_{b,m}^{(2)}}'\,\rho_p\,\rho_n
-4\,m\,b_0\,
\left(f_{b,p}^{(2)}\,\rho_p^2
+f_{b,n}^{(2)}\,\rho_n^2
+f_{b,m}^{(2)}\,\rho_p\,\rho_n\right)
\nonumber
\\
&&
+\frac{m\,(1-3\,w_p)^2\,{\rm e}^{6\,m\,b_0\,y}}
{24\,b_0^2\,\omega^4\,\left({\rm e}^{m\,b_0}-1\right)}\,
\left[{\rm e}^{2\,m\,b_0\,y}
\left(6\,m^2+b_0^2\,k^2\,\omega^2\right)
-b_0^2\,k^2\,\omega^2\,{\rm e}^{m\,b_0}\right]\,
\rho_p^2
\nonumber
\\
&&
+\frac{m\,(1-3\,w_n)^2\,{\rm e}^{2\,m\,b_0\,(3y-1)}}
{24\,b_0^2\,\omega^4\,\left({\rm e}^{m\,b_0}-1\right)}\,
\left[{\rm e}^{2\,m\,b_0\,y}
\left(6\,m^2+b_0^2\,k^2\,\omega^2\right)
-b_0^2\,k^2\,\omega^2\right]\,\rho_n^2
\nonumber
\\
&&
+\frac{m\,(1-3\,w_p)(1-3\,w_n)\,{\rm e}^{m\,b_0\,(6y-1)}}
{24\,b_0^2\,\omega^4\,\left({\rm e}^{m\,b_0}-1\right)}\,
\left[2\,{\rm e}^{2\,m\,b_0\,y}
\left(6\,m^2+b_0^2\,k^2\,\omega^2\right)
-b_0^2\,k^2\,\omega^2\left(1+{\rm e}^{m\,b_0}\right)
\right]\,\rho_p\,\rho_n=0
\ ,
\ee
which contains the first order parameters previously
determined. 
The solutions are 
\begin{subequations}
\be
\!\!\!\!
f_{b,p}^{(2)}\!\!&=&\!\!
{\rm e}^{4\,m\,b_0\,y}\left\{
b_p^{(2)}
-\frac{\left(1-3w_p\right)^2\,{\rm e}^{4\,m\,b_0\,y}}
{96\,b_0^3\,\omega^4\left({\rm e}^{m\,b_0}-1\right)^2}
\left[
6\,m^2+b_0^2\,k^2\,\omega^2
-2\,b_0^2\,k^2\,\omega^2\,{\rm e}^{m\,b_0\,(2y-3)}\right]
\right\}
\label{fb2_p}
\\
\nonumber
\\
\!\!\!\!
f_{b,n}^{(2)}\!\!&=&\!\!
{\rm e}^{4\,m\,b_0\,y}\left\{
b_n^{(2)}
-\frac{\left(1-3w_n\right)^2\,{\rm e}^{6\,m\,b_0\,(y-1)}}
{96\,b_0^3\,\omega^4\left({\rm e}^{m\,b_0}-1\right)^2}\,
\left[{\rm e}^{2\,m\,b_0\,y}
\left(6\,m^2+b_0^2\,k^2\,\omega^2\right)
-2\,b_0^2\,k^2\,\omega^2\right]
\right\}
\label{fb2_n}
\\
\nonumber
\\
\!\!\!\!
f_{b,m}^{(2)}\!\!&=&\!\!
{\rm e}^{4\,m\,b_0\,y}\,\left\{
b_m^{(2)}
-\frac{\left(1-3w_p\right)\left(1-3w_n\right)\,
{\rm e}^{m\,b_0\,(6y-5)}}
{48\,b_0^3\,\omega^4\left({\rm e}^{m\,b_0}-1\right)^2}\,
\left[{\rm e}^{2\,m\,b_0\,y}
\left(6\,m^2+b_0^2\,k^2\,\omega^2\right)
-b_0^2\,k^2\,\omega^2\left(1
+{\rm e}^{m\,b_0}\right)\right]
\right\}
\ ,
\label{fb2_m}
\ee
\end{subequations}
\end{widetext}
where
\be
b_p^{(2)}
\ ,
\quad
b_n^{(2)}
\ ,
\quad
b_m^{(2)}
\ ,
\ee
are integration constants to be determined from
the junction conditions. 
Once we plug the $f_{b,i}^{(2)}$'s into the second
order terms in Eq.~(\ref{F00}), we get the equations
for the $f_{a,i}^{(2)}$'s, which are not displayed
for the sake of brevity. 
Finally, by solving $G_{44}=k^2\,T_{44}$ one obtains
the corrections $f_{n,i}^{(2)}$'s.
The results will contain six integration constants
from the solutions of the first order differential
equations for $f_{a,i}^{(2)}$ and $f_{n,i}^{(2)}$,
\be
c_{a,i}^{(2)}
\ ,
\quad
c_{n,i}^{(2)}
\ ,
\label{2ordIntConst}
\ee
and nine parameters related to the radion, the
four-dimensional Hubble parameter and $c(t)$
respectively,
\be
h_{h,i}^{(2)}
\ ,
\quad
b_{i}^{(2)}
\ ,
\quad c_i^{(2)}
\ ,
\label{2ordCoeff}
\ee
where $i$ runs over $p$, $n$ and $m$ for second order
quantities. 
Analogously to the first order case, one can fix the
coefficients (\ref{2ordCoeff}) by imposing the junction
conditions, which form a system of nine independent
equations.
Nonetheless, one is again not able to fix the
constants (\ref{2ordIntConst}). 
Here are the solutions obtained at the end of the
calculations described above:
\begin{widetext}
\begin{subequations}
\be
f_{a,p}^{(2)}&=&
\frac{1}{4608\,m^2\,b_0^4\,\omega^4\,
\left({\rm e}^{m\,b_0}-1\right)^2}\,
\left\{2\,{\rm e}^{4\,m\,b_0\,y}\,
\left[3\,m^2(1-3w_p)^2-8\,b_0^2\,k^2\,\omega^2\right]
\left(6\,m^2-b_0^2\,k^2\omega^2\right)
\right.
\nonumber
\\
&&
\left.
+8\,b_0^2\,k^2\,\omega^2\,
{\rm e}^{m\,b_0\,(6y+1)}
\left[11\,m^2\,(3w_p-1)+2\,b_0^2\,k^2\omega^2\right]
\right.
\nonumber
\\
&&
\left.
-4\,b_0^2\,k^2\,\omega^2\,
{\rm e}^{m\,b_0\,(2y+1)}\,
\left[3\,m^2\,(9\,w_p^2+18\,w_p-7)
+8\,b_0^2\,k^2\omega^2\right]
\right.
\nonumber
\\
&&
\left.
+4\,b_0^2\,k^2\omega^2\,{\rm e}^{2\,m\,b_0\,(y+1)}
\left[27\,m^2\,(3w_p^2+2w_p-1)+16\,b_0^2\,k^2\omega^2\right]
+\right.
\nonumber
\\
&&
\left.
+{\rm e}^{8\,m\,b_0\,y}
\left[45\,m^4(1-3w_p)^2+3\,m^2b_0^2k^2\omega^2(9w_p^2-42w_p+13)
-4b_0^4k^4\omega^4\right]
\right.
\nonumber
\\
&&
\left.
-2\,{\rm e}^{2\,m\,b_0\,(2y+1)}
\left[81m^4(9w_p^3+9w_p^2-w_p-1)+3m^2b_0^2k^2
\omega^2(27w_p^2+42w_p-1)+8b_0^4k^4\omega^4\right]
\right.
\nonumber
\\
&&
\left.
-2{\rm e}^{m\,b_0\,(4y+1)}
\left[81\,m^4(9w_p^3+9w_p^2-w_p-1)
-3\,m^2b_0^2k^2\omega^2(9w_p^2+6w_p+29)
+16\,b_0^4k^4\omega^4\right]\right\}
+c_{a,p}^{(2)}
\ee
\be
f_{a,n}^{(2)}&=&
\frac{{\rm e}^{2mb_0(y-1)}}
{4608m^2b_0^4\omega^4\left({\rm e}^{mb_0}-1\right)^2}
\left\{2{\rm e}^{2mb_0(y+1)}
\left[3m^2(1-3w_n)^2-8b_0^2k^2\omega^2\right]
\left(6m^2-b_0^2k^2\omega^2\right)
\right.
\nonumber
\\
&&
\left.
+8b_0^2k^2\omega^2{\rm e}^{4mb_0y}
\left[11m^2(3w_n-1)+2b_0^2k^2\omega^2\right]
-4b_0^2k^2\omega^2{\rm e}^{2mb_0}
\left[3m^2(9w_n^2+18w_n-7)+8b_0^2k^2\omega^2\right]
\right.
\nonumber
\\
&&
\left.
+4b_0^2k^2\omega^2{\rm e}^{mb_0}
\left[27m^2(3w_n^2+2w_n-1)+16b_0^2k^2\omega^2\right]
\right.
\nonumber
\\
&&
\left.
+{\rm e}^{6mb_0y}\left[45m^4(1-3w_n)^2+3m^2b_0^2k^2
\omega^2(9w_n^2-42w_n+13)-4b_0^4k^4\omega^4\right]
\right.
\nonumber
\\
&&
\left.
-2{\rm e}^{2mb_0y}
\left[81m^4(9w_n^3+9w_n^2-w_n-1)
+3m^2b_0^2k^2\omega^2(27w_n^2+42w_n-1)+8b_0^4k^4
\omega^4\right]
\right.
\nonumber
\\
&&
\left.
-2{\rm e}^{mb_0(2y+1)}
\left[81m^4(9w_n^3+9w_n^2-w_n-1)
-3m^2b_0^2k^2\omega^2(9w_n^2+6w_n+29)+16b_0^4k^4
\omega^4\right]\right\}+c_{a,n}^{(2)}
\ee
\be
f_{a,m}^{(2)}&=&
\frac{{\rm e}^{mb_0(2y-1)}}
{2304m^2b_0^4\omega^4\left({\rm e}^{mb_0}-1\right)^2}
\left\{4b_0^2k^2\omega^2{\rm e}^{4mb_0y}
\left[m^2(3w_p-1)(6w_p-6w_n+11)
+2b_0^2k^2\omega^2\right]
\right.
\nonumber
\\
&&
\left.
+4b_0^2k^2\omega^2{\rm e}^{mb_0(4y+1)}
\left[m^2(3w_n-1)(6w_n-6w_p+11)
+2b_0^2k^2\omega^2\right]
\right.
\nonumber
\\
&&
\left.
+4b_0^2k^2\omega^2{\rm e}^{mb_0}
\left[3m^2(18w_p^2-9w_pw_n+9w_p-9w_n-1)
+4b_0^2k^2\omega^2\right]
\right.
\nonumber
\\
&&
\left.
+4b_0^2k^2\omega^2{\rm e}^{2mb_0}
\left[3m^2(18w_n^2-9w_pw_n+9w_n-9w_p-1)
+4b_0^2k^2\omega^2\right]
\right.
\nonumber
\\
&&
\left.
+{\rm e}^{6mb_0y}\left[45m^4(3w_p-1)(3w_n-1)+3m^2b_0^2k^2
\omega^2(9w_pw_n-21w_p-21w_n+13)-4b_0^4k^4\omega^4\right]
\right.
\nonumber
\\
&&
\left.
-{\rm e}^{2mb_0y}
\left[9m^4(3w_p-1)(18w_p^2+9w_pw_n+27w_p+3w_n+11)
\right.
\right.
\nonumber
\\
&&
\left.
\left.
+36m^2b_0^2k^2\omega^2(3w_p^2-3w_pw_n+w_p-2w_n-1)
+8b_0^4k^4\omega^4\right]
\right.
\nonumber
\\
&&
\left.
-{\rm e}^{2mb_0(y+1)}
\left[9m^4(3w_n-1)(18w_n^2+9w_pw_n+27w_n+3w_n+11)
\right.
\right.
\nonumber
\\
&&
\left.
\left.
+36m^2b_0^2k^2\omega^2(3w_n^2-3w_pw_n+w_n-2w_p-1)
+8b_0^4k^4\omega^4\right]
\right.
\nonumber
\\
&&
\left.
-{\rm e}^{mb_0(2y+1)}
\left[81m^4(6w_p^3+6w_n^3+3w_p^2w_n+3w_pw_n^2
+7w_p^2+7w_n^2+4w_pw_n-w_p-w_n-2)
\right.
\right.
\nonumber
\\
&&
\left.
\left.
+6m^2b_0^2k^2\omega^2(18w_p^2+18w_n^2-9w_pw_n+21w_p+21w_n-1)
+16b_0^4k^4\omega^4\right]\right\}
+c_{a,m}^{(2)}
\ee
\be
f_{n,p}^{(2)}&=&
\frac{{\rm e}^{2mb_0y}}
{4608m^2b_0^4\omega^4\left({\rm e}^{mb_0}-1\right)^2}
\left\{-8b_0^2k^2\omega^2(3w_p+2)
{\rm e}^{3mb_0}\left[3m^2(3w_p-1)+2b_0^2k^2\omega^2(3w_p+2)\right]
\right.
\nonumber
\\
&&
\left.
-8b_0^2k^2\omega^2{\rm e}^{mb_0(4y+1)}
\left[m^2(81w_p^3+180w_p^2+24w_p-31)
+2b_0^2k^2\omega^2(6w_p+5)\right]
\right.
\nonumber
\\
&&
\left.
+2{\rm e}^{2mb_0y}
\left(6m^2-b_0^2k^2\omega^2\right)
\left[3m^2(3w_p-1)^2-4b_0^2k^2\omega^2(9w_p^2+6w_p-1)\right]
\right.
\nonumber
\\
&&
\left.
-4b_0^2k^2\omega^2{\rm e}^{mb_0}
\left[-3m^2(63w_p^2+30w_p-17)
+4b_0^2k^2\omega^2(9w_p^2+6w_p-1)\right]
\right.
\nonumber
\\
&&
\left.
+4b_0^2k^2\omega^2\left[-27m^2(18w_p^3+27w_p^2+4w_p-5)
+8b_0^2k^2\omega^2(9w_p^2+6w_p-1)\right]
\right.
\nonumber
\\
&&
\left.
+{\rm e}^{6mb_0y}\left[45m^4(3w_p-1)^2
+3m^2b_0^2k^2\omega^2(117w_p^2+30w_p-23)
+4b_0^4k^4\omega^4(6w_p+5)\right]
\right.
\nonumber
\\
&&
\left.
-2{\rm e}^{mb_0(2y+1)}
\left[81m^4(9w_p^3+9w_p^2-w_p-1)
-3m^2b_0^2k^2\omega^2(162w_p^3+279w_p^2+60w_p-25)
\right.
\right.
\nonumber
\\
&&
\left.
\left.
+8b_0^4k^4\omega^4(9w_p^2+6w_p-1)\right]
+2{\rm e}^{2mb_0(y+1)}
\left[-9m^4(81w_p^3+90w_p^2-15w_p-8)
\right.
\right.
\nonumber
\\
&&
\left.
\left.
+3m^2b_0^2k^2\omega^2(162w_p^3+243w_p^2+12w_p-53)
+4b_0^4k^4\omega^4(9w_p^2+24w_p+14)\right]\right\}
+c_{n,p}^{(2)}
\ee
\be
f_{n,n}^{(2)}&=&
\frac{{\rm e}^{2mb_0(y-1)}}
{4608m^2b_0^4\omega^4\left({\rm e}^{mb_0}-1\right)^2}
\left\{-4b_0^2k^2\omega^2(6w_n+5)
{\rm e}^{mb_0}\left[27m^2(3w_n^2+2w_n-1)
+8b_0^2k^2\omega^2\right]
\right.
\nonumber
\\
&&
\left.
-8b_0^2k^2\omega^2{\rm e}^{4mb_0y}
\left[m^2(81w_n^3+180w_n^2+24w_n-31)
+2b_0^2k^2\omega^2(6w_n+5)\right]
\right.
\nonumber
\\
&&
\left.
+4b_0^2k^2\omega^2{\rm e}^{2mb_0}
\left[3m^2(45w_n^2+24w_n-13)
+4b_0^2k^2\omega^2(6w_n+5)\right]
\right.
\nonumber
\\
&&
\left.
-2{\rm e}^{mb_0(2y+1)}
\left[81m^4(9w_n^3+9w_n^2-w_n-1)
-3m^2b_0^2k^2\omega^2(162w_n^3+243w_n^2+48w_n-17)
\right.
\right.
\nonumber
\\
&&
\left.
\left.
-8b_0^4k^4\omega^4(6w_n+5)\right]
+2{\rm e}^{2mb_0(y+1)}\left[9m^4(3w_n-1)^2
\right.
\right.
\nonumber
\\
&&
\left.
\left.
-3m^2b_0^2k^2\omega^2(45w_n^2+30w_n+1)
-4b_0^4k^4\omega^4(6w_n+5)\right]
+{\rm e}^{6mb_0y}\left[45m^4(3w_n-1)^2
\right.
\right.
\nonumber
\\
&&
\left.
\left.
+3m^2b_0^2k^2\omega^2(117w_n^2+30w_n-23)
+4b_0^4k^4\omega^4(6w_n+5)\right]
+2{\rm e}^{2mb_0y}\left[-81m^4(9w_n^3+9w_n^2-w_n-1)
\right.
\right.
\nonumber
\\
&&
\left.
\left.
+3m^2b_0^2k^2\omega^2(162w_n^3+243w_n^2+12w_n-53)
+8b_0^4k^4\omega^4(6w_n+5)\right]\right\}
+c_{n,n}^{(2)}
\ee
\be
f_{n,m}^{(2)}&=&
\frac{{\rm e}^{-mb_0}}{2304m^2b_0^4\omega^4
\left({\rm e}^{mb_0}-1\right)^2}
\left\{4b_0^2k^2\omega^2(3w_p+2)
{\rm e}^{mb_0(2y+3)}\left[3m^2(1-3w_n)
+2b_0^2k^2\omega^2(3w_n+2)\right]
\right.
\nonumber
\\
&&
\left.
-4b_0^2k^2\omega^2{\rm e}^{6mb_0y}
\left[m^2(3w_p-1)(18w_p^2+9w_pw_n+45w_p+24w_n+31)
+2b_0^2k^2\omega^2(3w_p+3w_n+5)\right]
\right.
\nonumber
\\
&&
\left.
-4b_0^2k^2\omega^2{\rm e}^{mb_0(6y+1)}
\left[m^2(3w_n-1)(18w_n^2+9w_pw_n+45w_n+24w_p+31)
+2b_0^2k^2\omega^2(3w_p+3w_n+5)\right]
\right.
\nonumber
\\
&&
\left.
-4b_0^2k^2\omega^2{\rm e}^{2mb_0(y+1)}
\left[3m^2(3w_n+2)(18w_n^2+9w_pw_n-12w_p+27w_n-8)
\right.
\right.
\nonumber
\\
&&
\left.
\left.
+2b_0^2k^2\omega^2(18w_pw_n+15w_p+15w_n+13)\right]
-4b_0^2k^2\omega^2{\rm e}^{mb_0(2y+1)}
\left[-2b_0^2k^2\omega^2(9w_pw_n+3w_p+3w_n-1)
\right.
\right.
\nonumber
\\
&&
\left.
\left.
+3m^2(3w_p+2)(18w_p^2+9w_pw_n+27w_p-15w_n-7)\right]
+{\rm e}^{8mb_0y}\left[45m^4(3w_p-1)(3w_n-1)
\right.
\right.
\nonumber
\\
&&
\left.
\left.
+3m^2b_0^2k^2\omega^2(117w_pw_n+15w_p+15w_n-23)
+4b_0^4k^4\omega^4(3w_p+3w_n+5)\right]
\right.
\nonumber
\\
&&
\left.
+{\rm e}^{mb_0(4y+1)}
\left[-81m^4(6w_p^3+6w_n^3+3w_p^2w_n
+3w_pw_n^2+7w_p^2+7w_n^2+4w_pw_n-w_p-w_n-2)
\right.
\right.
\nonumber
\\
&&
\left.
\left.
+6m^2b_0^2k^2\omega^2(54w_p^3+54w_n^3+27w_p^2w_n
+27w_pw_n^2+117w_p^2+117w_n^2-9w_pw_n-6w_p+12w_n-49)
\right.
\right.
\nonumber
\\
&&
\left.
\left.
+8b_0^4k^4\omega^4(9w_pw_n+12w_p+12w_n+14)\right]
+{\rm e}^{4mb_0y}
\left[-9m^4(3w_p-1)(18w_p^2+9w_pw_n+27w_p+3w_n+11)
\right.
\right.
\nonumber
\\
&&
\left.
\left.
+54m^2b_0^2k^2\omega^2
(6w_p^3+3w_p^2w_n+13w_p^2-2w_pw_n+4w_p-3w_n-1)
-4b_0^4k^4\omega^4(9w_pw_n+3w_p+3w_n-1)\right]
\right.
\nonumber
\\
&&
\left.
+{\rm e}^{2mb_0(2y+1)}
\left[-81m^4(w_p+2w_n+1)(3w_n^2+2w_n-1)
\right.
\right.
\nonumber
\\
&&
\left.
\left.
+6m^2b_0^2k^2\omega^2(54w_n^3+27w_pw_n^2+117w_n^2-15w_p+30w_n-13)
\right.
\right.
\nonumber
\\
&&
\left.
\left.
-4b_0^4k^4\omega^4(9w_pw_n+3w_p+3w_n-1)\right]\right\}
+c_{n,m}^{(2)}
\ .
\ee
\end{subequations}
\end{widetext}
\section{Second order cosmology}
\label{IV}
Let us now look at the physical consequences of the above
results as applied to two different cosmological scenarios. 
The Friedmann and acceleration equations will be showed and
their phenomenology investigated in different regimes of
$\omega$ and for different matter equations of state on
the branes. 
\subsection{RS~I}
We begin from the case with two branes~\footnote{The unperturbed
brane distance is taken to be finite and equal to $b_0/2$.}
for which we shall study how cosmology would be described
by observers on the negative brane.
In order to achieve that, one can use the corresponding
proper time by fixing the $c_{n,i}^{(1)}$'s as in
Eq.~(\ref{cn1negtime}), choosing
\begin{widetext}
\begin{subequations}
\be
c_{n,p}^{(2)}&=&
\frac{{\rm e}^{2mb_0}}
{4608m^2b_0^4\omega^4\left({\rm e}^{mb_0}-1\right)^2}
\left\{9m^4(3w_p-1)\left[3{\rm e}^{2mb_0}(3w_p^2+4w_p+1)
+18{\rm e}^{mb_0}(3w_p^2+4w_p+1)-12w_p+4\right]
\right.
\nonumber
\\
&&
\left.
-m^2b_0^2k^2\omega^2\left[{\rm e}^{2mb_0}
(324w_p^3+153w_p^2-102w_p-91)-6{\rm e}^{mb_0}
(162w_p^3+207w_p^2+12w_p-65)
\right.
\right.
\nonumber
\\
&&
\left.
\left.
+54(5w_p^2+2w_p-3)\right]+4b_0^4k^4\omega^4
\left[3{\rm e}^{2mb_0}
(6w_p^2+6w_p+1)+2\left(1-2{\rm e}^{mb_0}\right)
(9w_p^2+6w_p-1)\right]\right\}
\label{cnp2I}
\ee
\be
c_{n,n}^{(2)}&=&
\frac{1}{4608m^2b_0^4\omega^4\left({\rm e}^{mb_0}-1\right)^2}
\left\{9m^4(3w_n-1)\left[-7{\rm e}^{2mb_0}(3w_n-1)
+18\left({\rm e}^{mb_0}+1\right)(3w_n^2+4w_n+1)\right]
\right.
\nonumber
\\
&&
\left.
-m^2b_0^2k^2\omega^2\left[{\rm e}^{2mb_0}
(81w_n^2-90w_n-75)+2{\rm e}^{mb_0}(162w_n^3+279w_n^2+192w_n-5)
\right.
\right.
\nonumber
\\
&&
\left.
\left.
-6(162w_n^3+243w_n^2+60w_n-37)\right]
+4b_0^4k^4\omega^4\left({\rm e}^{mb_0}-2\right)^2
(6w_n+5)\right\}
\label{cnt2I}
\ee
\be
c_{n,m}^{(2)}&=&
\frac{{\rm e}^{mb_0}}{2304m^2b_0^4\omega^4
\left({\rm e}^{mb_0}-1\right)^2}
\left\{9m^4\left[{\rm e}^{2mb_0}(3w_n-1)
(18w_n^2+9w_pw_n-6w_p+27w_n+14)
\right.
\right.
\nonumber
\\
&&
\left.
\left.9{\rm e}^{mb_0}
(6w_p^3+6w_n^3+3w_p^2w_n+3w_pw_n^2
+7w_p^2+7w_n^2+4w_pw_n-w_p-w_n-2)+(3w_p-1)
(18w_p^2
\right.
\right.
\nonumber
\\
&&
\left.
\left.
+9w_pw_n+27w_p+3w_n+11)\right]
-m^2b_0^2k^2\omega^2\left[{\rm e}^{2mb_0}
(108w_n^3+54w_pw_n^2+234w_n^2-9w_pw_n+87w_p
\right.
\right.
\nonumber
\\
&&
\left.
\left.
-39w_n+1)+2{\rm e}^{mb_0}
(54w_p^3-162w_n^3+27w_p^2w_n-81w_pw_n^2
+117w_p^2-351w_n^2-45w_pw_n+30w_p
\right.
\right.
\nonumber
\\
&&
\left.
\left.
-96w_n+11)
-6(54w_p^3+27w_p^2w_n+117w_p^2-36w_pw_n+30w_p-33w_n-19)
\right]
\right.
\nonumber
\\
&&
\left.
-4b_0^4k^4\omega^4
\left[\left({\rm e}^{2mb_0}-2{\rm e}^{mb_0}\right)
(3w_p+2)(3w_n+2)+9w_pw_n+3w_p+3w_n-1\right]
\right\}
\label{cnm2I}
\ee
\end{subequations}
and finally rescaling $n(y,t)$ to satisfy the
condition $n(1/2,\tau)=1$.
The Friedmann equation we are interested in is given
by the second order expression of the Hubble parameter
at $y=1/2$ as a function of $\tau$,
\be
H^2(1/2,\tau)&=&
\frac{k^2m}{3\left({\rm e}^{mb_0}-1\right)}
\left(\rho_n+{\rm e}^{2mb_0}\rho_p\right)-(w_n+1)
\left(3w_n-1\right)
\frac{3m^2k^2\left({\rm e}^{mb_0}+1\right)}
{32b_0^2\omega^2\left({\rm e}^{mb_0}-1\right)^2}\,
\rho_n^2
\nonumber
\\
&&
-\left\{\frac{k^4}{36}
+\frac{m^2k^2\left({\rm e}^{mb_0}+1\right)}
{48b_0^2\omega^2\left({\rm e}^{mb_0}-1\right)^2}\,
\left[18w_p^2-9w_p(w_n-1)-9w_n-1+2{\rm e}^{mb_0}
(9w_n^2+9w_n-4)\right]\right\}
{\rm e}^{mb_0}\,\rho_p\,\rho_n
\nonumber
\\
&&
+\left\{\frac{k^4}{36}
+\frac{m^2k^2(3w_p-1)
\left({\rm e}^{mb_0}+1\right)}
{96b_0^2\omega^2\left({\rm e}^{mb_0}-1\right)^2}\,
\left[3w_p+7-4{\rm e}^{mb_0}(3w_p+4)\right]\right\}
{\rm e}^{2mb_0}\,
\rho_p^2
\ .
\label{2ordnegFrEq}
\ee
Note that the results~(\ref{cnp2I})-(\ref{2ordnegFrEq})
do not depend on the integration constants
$c_{a,i}^{(1)}$'s and $c_{a,i}^{(2)}$'s,
which reflects the fact that the three-dimensional spatial
curvature has been set to zero ab initio.
\par
The Friedmann equation contains coefficients up to
second order in the vacuum perturbations of both
branes.
First and second order contributions in Eq.~(\ref{2ordnegFrEq})
are consequences of the adiabatic regime of the five-dimensional
dynamics which determines the value of the integration constant
$\tilde c(t)$ in Eq.~(\ref{F00}).
The value of the latter is affected by the presence of matter
on both branes through the junction conditions and, for instance,
up to $\mathcal{O}(\epsilon^2)$ is given by  
\begin{subequations}
\be
\lim_{\omega\to\infty} c^{(1)}&=&
\frac{m\,k^2\,\left(\rho_p+{\rm e}^{mb_0}\rho_n\right)}
{3\,\left({\rm e}^{m\,b_0}-1\right)}
\\
\nonumber
\\
\lim_{\omega\to\infty} c^{(2)}&=&
\frac{k^4}{36}\,
\left[\frac{
{\rm e}^{2\,m\,b_0}+2\,{\rm e}^{m\,b_0}-1}
{\left({\rm e}^{m\,b_0}-1\right)^2}\,\rho_p^2
+\left(\frac{{\rm e}^{m\,b_0}+1}
{{\rm e}^{m\,b_0}-1}\right)^2
{\rm e}^{-m\,b_0}\,
\rho_p\rho_n
+\frac{{\rm e}^{-2\,m\,b_0}+2\,{\rm e}^{-m\,b_0}-1}
{\left({\rm e}^{m\,b_0}-1\right)^2}\,
\rho_n^2
\right]
\ ,
\ee
\end{subequations}
\end{widetext}
in the limit of infinite spring constant.
Furthermore, as previously noted, this effect is also
a consequence of the radion field potential acting as
a source in Eq.~(\ref{F00}).
One can expect to cancel some terms in Eq.~(\ref{2ordnegFrEq})
by arbitrarily increasing the spring constant of the effective
radion potential in order to decrease the radion shift 
from equilibrium.
This mechanism partially works as the first order contributions
to the bulk potential vanish along the time-time direction,
whereas only second order terms, which depend on the matter
equation of state, cancel when $\omega\to\infty$.
One is then left with
\be
\!\!\!\!\!\!
\lim_{\omega\to\infty}H^2(1/2,\tau)
\!\!&=&\!\!
\frac{m\,k^2\,\left(\rho_n+{\rm e}^{2\,m\,b_0}\rho_p\right)}
{3\,\left({\rm e}^{mb_0}-1\right)}
\nonumber
\\
\!\!&&\!\!
-\frac{k^4}{36}\,{\rm e}^{m\,b_0}\,\rho_p\,\rho_n
+\frac{k^4}{36}\,{\rm e}^{2\,m\,b_0}\,\rho_p^2 
\ ,
\label{FrEqIOmegainf}
\ee
which is analogous to what has been obtained
in Ref.~\cite{vinet}.
The matter on the positive tension brane appears at
second order with the role of some ``dark'' fluid
and acts as a sort of Brans-Dicke field which adiabatically
modifies the Newton constant perceived in the visible Universe.
This behavior is somehow inherited by the dynamics of the radion
which is known to modulate the strength of gravity on the visible
brane.
\par
Note that while first order coefficients are positive definite,
irrespective of the brane tension, the sign of second order ones
depends on the matter equation of state.
Differently enough from unstabilized brane cosmology,
this fact implies that leading order cosmological equations
have the correct behavior on both branes.
Letting the energy density $\rho_p\to 0$ in Eq.~(\ref{2ordnegFrEq}),
one obtains
\be
H^2(1/2,\tau)\!\!&=&\!\!
(w_n+1)\left(1-3w_n\right)
\frac{3\,m^2\,k^2\left({\rm e}^{m\,b_0}+1\right)}
{32\,b_0^2\omega^2\left({\rm e}^{m\,b_0}-1\right)^2}\,
\rho_n^2
\nonumber
\\
&&
+\frac{m\,k^2\,\rho_n}{3\,\left({\rm e}^{m\,b_0}-1\right)}
\ ,
\label{FrEqRhoPzero}
\ee
which has the usual first order solution for both radiation
and a cosmological constant on the visible brane.
A matter dominated Universe would otherwise generate
second order corrections.
\par
We now come to the equation for the acceleration,
which has the general form
\begin{widetext}
\be
\frac{\ddot a(1/2,\tau)}{a(1/2,\tau)}
\!\!&=&\!\!
\frac{m\,k^2}{6\left(1-{\rm e}^{m\,b_0}\right)}
\left[(3w_n+1)\rho_n
+{\rm e}^{2\,m\,b_0}(3w_p+1)\rho_p\right]
+(w_n+1)\left(3w_n-1\right)
\left(3w_n+2\right)
\frac{3m^2k^2\left({\rm e}^{m\,b_0}+1\right)}
{32b_0^2\omega^2\left({\rm e}^{mb_0}-1\right)^2}\rho_n^2
\nonumber
\\
&&
-\left\{\frac{k^4}{144}(3w_n+1)(3w_p+1)
+\frac{m^2k^2 }{96b_0^2\omega^2\left({\rm e}^{m\,b_0}-1\right)^2}
\left[{\rm e}^{2\,m\,b_0}(3w_n-1)(18w_n^2+9w_pw_n+15w_p+39w_n+23)
\right.
\right.
\nonumber
\\
&&
\left.
\phantom{\frac{A}{B}}
\left.
+9{\rm e}^{m\,b_0}
(6w_p^3+w_n^3+3w_p^2w_n+3w_pw_n^2+11w_p^2+11w_n^2+2w_pw_n-4)
\right.
\right.
\nonumber
\\
&&
\left.
\phantom{\frac{A}{B}}
\left.
+54w_p^3-27w_p^2w_n+99w_p^2-18w_pw_n+24w_p-13\right]
\right\}
{\rm e}^{m\,b_0}\,\rho_p\,\rho_n
\nonumber
\\
&&
+\left\{\frac{k^4}{144}(3w_p+1)^2
+\frac{m^2k^2(3w_p-1)\left({\rm e}^{m\,b_0}+1\right)}
{96b_0^2\omega^2\left({\rm e}^{m\,b_0}-1\right)^2}
\left[{\rm e}^{m\,b_0}(27w_p^2+54w_p+23)
-9w_p-5\right]\right\}
{\rm e}^{2\,m\,b_0}
\rho_p^2
\ ,
\label{2ordnegAccEq}
\ee
and, for $\omega\to\infty$, reduces to 
\be
\frac{\ddot a(1/2,\tau)}{a(1/2,\tau)}&=&
\frac{m\,k^2}{6\left(1-{\rm e}^{m\,b_0}\right)}
\left[(3w_n+1)\rho_n+{\rm e}^{2\,m\,b_0}(3w_p+1)\,
\rho_p\right]
\nonumber
\\
&&
-\frac{k^4}{144}(3w_n+1)(3w_p+1){\rm e}^{m\,b_0}\,
\rho_p\,\rho_n
+\frac{k^4}{144}(3w_p+1)^2{\rm e}^{2\,m\,b_0}\,
\rho_p^2
\ ,
\label{2ordnegAccEqOmegaInf}
\ee
which is again analogous to the result of Ref.~\cite{vinet}.
The coefficient of $\rho_p^2$ in Eq.~(\ref{2ordnegAccEqOmegaInf})
is positive or zero in this limit and provides an accelerating
contribution to the equation.
The coefficient of the mixed term has a positive value when
just one fluid has $w_i<-1/3$.
For $\rho_p\to 0$, one is further left with
\be
\frac{\ddot a(1/2,\tau)}{a(1/2,\tau)}=
\frac{m\,k^2\,(3w_n+1)}{6\left(1-{\rm e}^{mb_0}\right)}\,\rho_n
+(w_n+1)\left(3w_n-1\right)\left(3w_n+2\right)
\frac{3m^2k^2\left({\rm e}^{m\,b_0}+1\right)}
{32b_0^2\omega^2\left({\rm e}^{m\,b_0}-1\right)^2}\,
\rho_n^2
\ ,
\label{2ordnegAccEqRhoPzero}
\ee
\end{widetext}
where the second order contribution is inversely proportional
to $\omega$ and vanishes for radiation and a cosmological constant.
This peculiarity leads to the standard cosmological evolution up
to $\mathcal{O}(\rho_n^2)$ until the matter dominated era.
\par
A singularity in the lapse function $n(y,t)$ for
$b_0\to\infty$ prevents us from analyzing the correct 
limit when the distance between the branes becomes infinite,
we shall thus comment on this problem in the next subsection. 
On setting $\rho_p\to 0$, Eq.~(\ref{FrEqIOmegainf})
admits the finite but trivial limit
\begin{subequations}
\be
\lim _{b_0\to \infty}H^2(1/2,\tau)=0
\ ,
\ee
in which one also has
\be
\lim _{b_0\to \infty}
\frac{\ddot a(1/2,\tau)}{a(1/2,\tau)}
=0
\ .
\ee
\end{subequations}
This result is due to the reduced strength of
the gravitational interaction at infinity.
\subsection{RS~II}
On can think of the RS~II model as the limit of RS~I in which
the distance between the two branes becomes infinite,
thus one expects that only $\rho_p$ contributes in this limit. 
The cosmological proper time is now the one on the Planck brane
and is recovered upon choosing
\begin{widetext}
\begin{subequations}
\be
c_{n,p}^{(2)}&=&
\frac{1}{4608m^2b_0^4\omega^4\left({\rm e}^{mb_0}-1\right)^2}
\left\{9m^4(3w_p-1)\left[18{\rm e}^{2mb_0}(3w_p^2+4w_p+1)
+18{\rm e}^{mb_0}(3w_p^2+4w_p+1)
\right.
\right.
\nonumber
\\
&&
\left.
\left.
-21w_p+7\right]+m^2b_0^2k^2\omega^2\left[6{\rm e}^{2mb_0}(162w_p^3+243w_p^2+60w_p-37)
\right.
\right.
\nonumber
\\
&&
\left.
\left.
-2{\rm e}^{mb_0}(162w_p^3+279w_p^2+192w_p-5)-81w_p^2+90w_p+75\right]
\right.
\nonumber
\\
&&
\left.
+4b_0^4k^4\omega^4\left(1-2{\rm e}^{mb_0}\right)^2(6w_p+5)\right\}
\ee
\be
c_{n,n}^{(2)}&=&
\frac{{\rm e}^{-2mb_0}}
{4608m^2b_0^4\omega^4\left({\rm e}^{mb_0}-1\right)^2}
\left\{9m^4(3w_n-1)\left[4{\rm e}^{2mb_0}(1-3w_n)
+18{\rm e}^{mb_0}(3w_n^2+4w_n+1)
\right.
\right.
\nonumber
\\
&&
\left.
\left.
+54w_n^2+63w_n+21\right]
-m^2b_0^2k^2\omega^2\left[54{\rm e}^{2mb_0}(5w_n^2+2w_n-3)
\right.
\right.
\nonumber
\\
&&
\left.
\left.
-6{\rm e}^{mb_0}(162w_n^3+207w_n^2+12w_n-65)
+324w_n^3+153w_n^2-102w_n-91\right]
\right.
\nonumber
\\
&&
\left.
+4b_0^4k^4\omega^4\left[2\left({\rm e}^{2mb_0}
-2{\rm e}^{mb_0}\right)(9w_n^2+6w_n-1)
+3(6w_n^2+6w_n+1)\right]\right\}
\ee
\be
c_{n,m}^{(2)}&=&
\frac{{\rm e}^{mb_0}}
{2304m^2b_0^4\omega^4\left({\rm e}^{mb_0}-1\right)^2}
\left\{9m^4\left[{\rm e}^{2mb_0}(3w_n-1)(18w_n^2+9w_pw_n+3w_p+27w_n+11)
\right.
\right.
\nonumber
\\
&&
\left.
\left.
+9{\rm e}^{mb_0}
(6w_p^3+6w_n^3+3w_p^2w_n+3w_pw_n^2
+7w_p^2+7w_n^2+4w_pw_n-w_p-w_n-2)
\right.
\right.
\nonumber
\\
&&
\left.
\left.
(3w_p-1)(18w_p^2+9w_pw_n+27w_p-6w_n+14)\right]
+m^2b_0^2k^2\omega^2\left[6{\rm e}^{2mb_0}
(54w_n^2++27w_pw_n^2
\right.
\right.
\nonumber
\\
&&
\left.
\left.
+117w_n^2-36w_pw_n-33w_p+30w_n-19)
+2{\rm e}^{mb_0}(162w_p^3-54w_n^3+81w_p^2w_n-27w_pw_n^2
\right.
\right.
\nonumber
\\
&&
\left.
\left.
+351w_p^2-117w_n^2+96w_p-30w_n-11)
-108w_p^3-54w_p^2w_n-234w_p^2+9w_pw_n+39w_p-87w_n-1\right]
\right.
\nonumber
\\
&&
\left.
-4b_0^4k^4\omega^4\left[{\rm e}^{2mb_0}(9w_pw_n+3w_p+3w_n-1)
+\left(1-2{\rm e}^{mb_0}\right)(3w_p+2)(3w_n+2)\right]
\right\}
\ee
\end{subequations}
together with Eq.~(\ref{cn1postime}).
The Friedmann equation, for a finite $b_0$, reduces to
\be
H^2(0,t)&=&
\frac{mk^2}{3\left({e}^{m\,b_0}-1\right)}\,
\left({\rm e}^{m\,b_0}\,\rho_p
+{\rm e}^{-m\,b_0}\,\rho_n\right)
+(w_p+1)\left(1-3w_p\right)
\frac{3m^2k^2{\rm e}^{m\,b_0}
\left({\rm e}^{m\,b_0}+1\right)}
{32b_0^2\omega^2\left({\rm e}^{m\,b_0}-1\right)^2}\,
\rho_p^2
\nonumber
\\
&&
-\left\{\frac{k^4}{36}+\frac{m^2k^2\left({\rm e}^{m\,b_0}+1\right)
{\rm e}^{m\,b_0}
(18w_n^2-9w_pw_n-9w_p+9w_n-1)(9w_p^2+9w_p-4)}
{24b_0^2\omega^2\left({\rm e}^{mb_0}-1\right)^2}
\right\}{\rm e}^{-m\,b_0}\,\rho_p\,\rho_n
\nonumber
\\
&&
+\left\{\frac{k^4}{36}
+\frac{m^2k^2\left({\rm e}^{m\,b_0}+1\right)(3w_n-1)
\left[{\rm e}^{m\,b_0}(3w_n+7)-4(3w_n+4)\right]}
{96b_0^2\omega^2\left({\rm e}^{m\,b_0}-1\right)^2}
\right\}{\rm e}^{-2\,m\,b_0}\,
\rho_n^2
\ ,
\ee
\end{widetext}
which is similar to Eq.~(\ref{2ordnegFrEq}) but has
the finite, non trivial limit
\be
\lim_{b_0\to\infty}H^2(0,t)=\frac{m\,k^2}{3}\,\rho_p
\ .
\label{infb0H2}
\ee
This is precisely the standard Friedmann equation one
has in four-dimensional cosmology. 
Furthermore, the equation for the acceleration to second
order in terms of the time on the positive tension brane is
given by
\begin{widetext}
\be
\frac{\ddot a(0,t)}{a(0,t)}\!\!&=&\!\!
\frac{m\,k^2\,{\rm e}^{m\,b_0}}{6\left(1-{\rm e}^{m\,b_0}\right)}
\left[(3w_p+1)
\rho_p+(3w_n-1){\rm e}^{-2\,m\,b_0}\rho_n\right]
+\left(w_p+1\right)\left(3w_p-1\right)\left(3w_p+2\right)
\frac{3m^2k^2{\rm e}^{m\,b_0}
\left({\rm e}^{m\,b_0}+1\right)}
{32b_0^2\omega^2\left({\rm e}^{m\,b_0}-1\right)^2}\rho_p^2
\nonumber
\\
&&
-\left\{\frac{k^4}{144}(3w_p+1)(3w_n+1)-
\frac{m^2k^2}{96b_0^2\omega^2\left({\rm e}^{mb_0}-1\right)^2}
\left[{\rm e}^{2mb_0}(54w_n^3+27w_pw_n^2+99w_n^2-18w_pw_n
\right.
\right.
\nonumber
\\
&&
\left.
\phantom{\frac{A}{B}}
\left.
-21w_p+24w_n-13)+9{\rm e}^{mb_0}
(6w_p^3+6w_n^3+3w_p^2w_n+3w_pw_n^2+11w_p^2+11w_n^2+2w_pw_n
\right.
\right.
\nonumber
\\
&&
\left.
\phantom{\frac{A}{B}}
\left.
+w_p+w_n-4)+(3w_p-1)(18w_p^2+9w_pw_n+39w_p+15w_n+23)
\right]\right\}
{\rm e}^{-mb_0}
\rho_p\rho_n
\nonumber
\\
&&
+\left\{\frac{k^4}{144}(3w_n+1)^2
-\frac{m^2k^2\left({\rm e}^{mb_0}\right)(3w_n-1)}
{96b_0^2\omega^2\left({\rm e}^{mb_0}-1\right)^2}
\left[{\rm e}^{mb_0}(9w_n+5)-27w_n^2+54w_n+23\right]
\right\}{\rm e}^{-2mb_0}\rho_n^2
\ ,
\ee
\end{widetext}
and, compatibly with Eq.~(\ref{infb0H2}), yields
\be
\lim_{b_0\to\infty}
\frac{\ddot a(0,t)}{a(0,t)}=
-\frac{m\,k^2}{6}(3w_p+1)\,\rho_p
\ ,
\ee
in the limit of infinite brane distance.
Thus second order effects, typical of 
brane cosmology, become more and more negligible
when the distance between the branes grows,
which consequently leads to unobservable deviations
from standard four-dimensional General Relativity.
\section{Approximation analysis}
\label{V}
The results obtained so far hold with the assumption that
$\rho_i\ll M^4$ where $M$ is, in general, the natural mass
scale of the model.
By taking all the mass parameters to such a natural scale,
one expects the solutions to second order provide a good
approximation for Eq.~(\ref{Ein1}). 
In this regime, however, second order effects are certainly
sub-leading and thus insufficient to significantly alter
first order behavior.
We shall hence present below a numerical analysis of the
validity of our approximate solutions in the attempt to
extend the range of the parameters in which ours
results hold valid and widen the conclusions one can draw
from second order expressions. 
In particular, we are interested in possible deviations
from standard cosmological equations due to terms of
$\mathcal{O}(\epsilon^2)$ in the first and third of
Eqs.~(\ref{Ein}).
Note that all numerical results will be obtained by
setting the expansion parameter $\epsilon=1$ as previously
prescribed.
\par
In order to test our approximations, since Eqs.~(\ref{Ein1})
are not analytically solvable, we substitute the second order
solutions into the exact Einstein equations to obtain an estimate
of their non vanishing remainders which we then compare with the
leading contributions (satisfying the corresponding approximate
equations).
Since the time dependence is contained in $\rho_i$ in our
expansions (\ref{genexp})-(\ref{cexp}), it is also convenient
to trade the time for $\rho$.
We thus divide Eq.~(\ref{F00}) into the following five terms
\begin{subequations}
\be
\eta_1(y,\rho)&\equiv&
\frac{1}{n^2(y,t)}\,\frac{\dot a^2(y,t)}{a^2(y,t)}
\label{F00terms1}
\\
\nonumber
\\
\eta_2(y,\rho)&\equiv&
-\frac{1}{b^2(y,t)}\,
\left(\frac{a'(y,t)}{a(y,t)}\right)^2
\label{F00terms2}
\\
\nonumber
\\
\eta_3(y,\rho)&\equiv&
-\frac{k^2}{6}(\Lambda+U)
\label{F00terms3}
\\
\nonumber
\\
\eta_4(y,\rho)&\equiv&
\frac{k^2}{6\,a^4(y,t)}\,\int^y
a^4\,\left(T^0_{\ 0}\right)'\,\d x
\label{F00terms4}
\\
\nonumber
\\
\eta_5(y,\rho)&\equiv&
-\frac{\tilde c(t)}{a^4(y,t)}
\ ,
\label{F00terms5}
\ee
\end{subequations}
in which $t=t(\rho)$ in the right~hand~sides is understood
as the time when $\rho=\rho_n$ ($\rho_p$ will be chosen either
equal to $\rho_n$ or zero).
The sum,
\be
R_\eta(y,\rho)\equiv
\sum_l\eta_l
\ ,
\ee
evaluated on the second order solutions yields
$R_\eta(y,\rho)=\mathcal{O}(\epsilon^3)$ as a measure
of the corresponding error.
Similarly, the third of Eqs.~(\ref{Ein1}) may be written
as the sum of the following six terms
\begin{figure*}[ht]
{\includegraphics[width=3in]{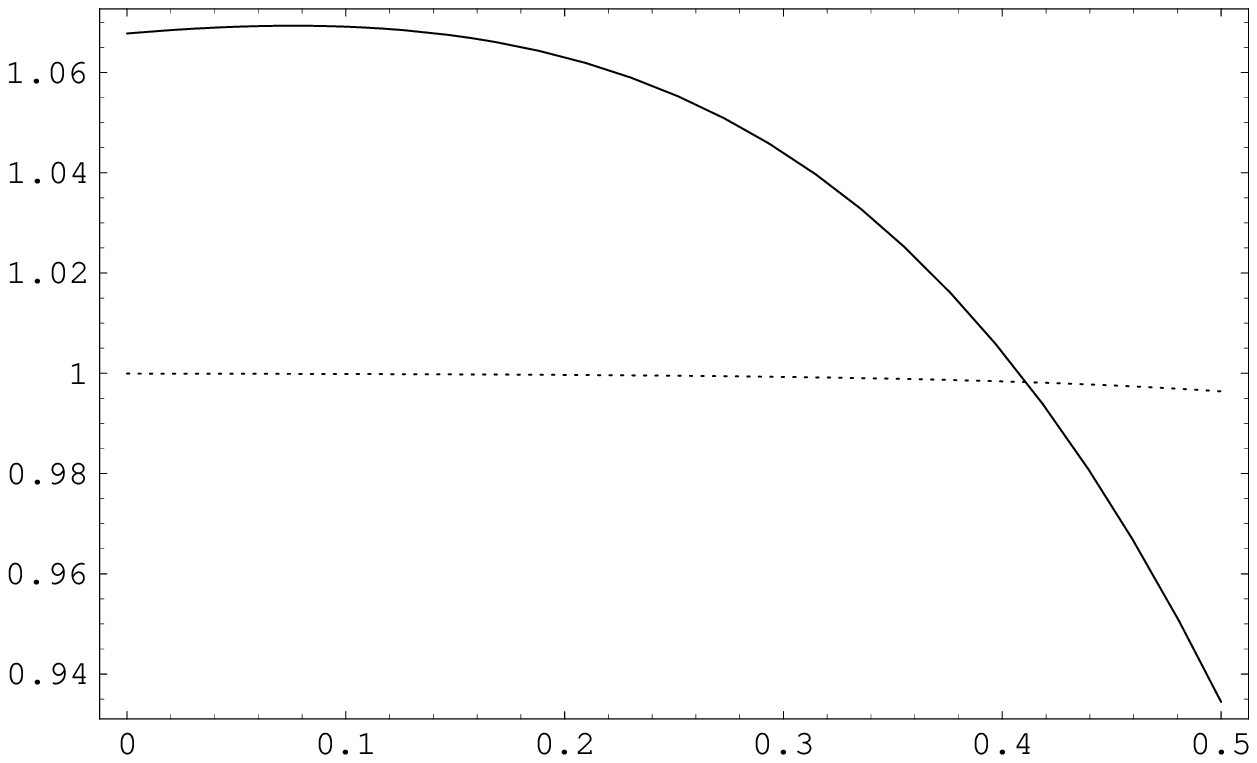}
\hspace{0,52 cm}
\includegraphics[width=3in]{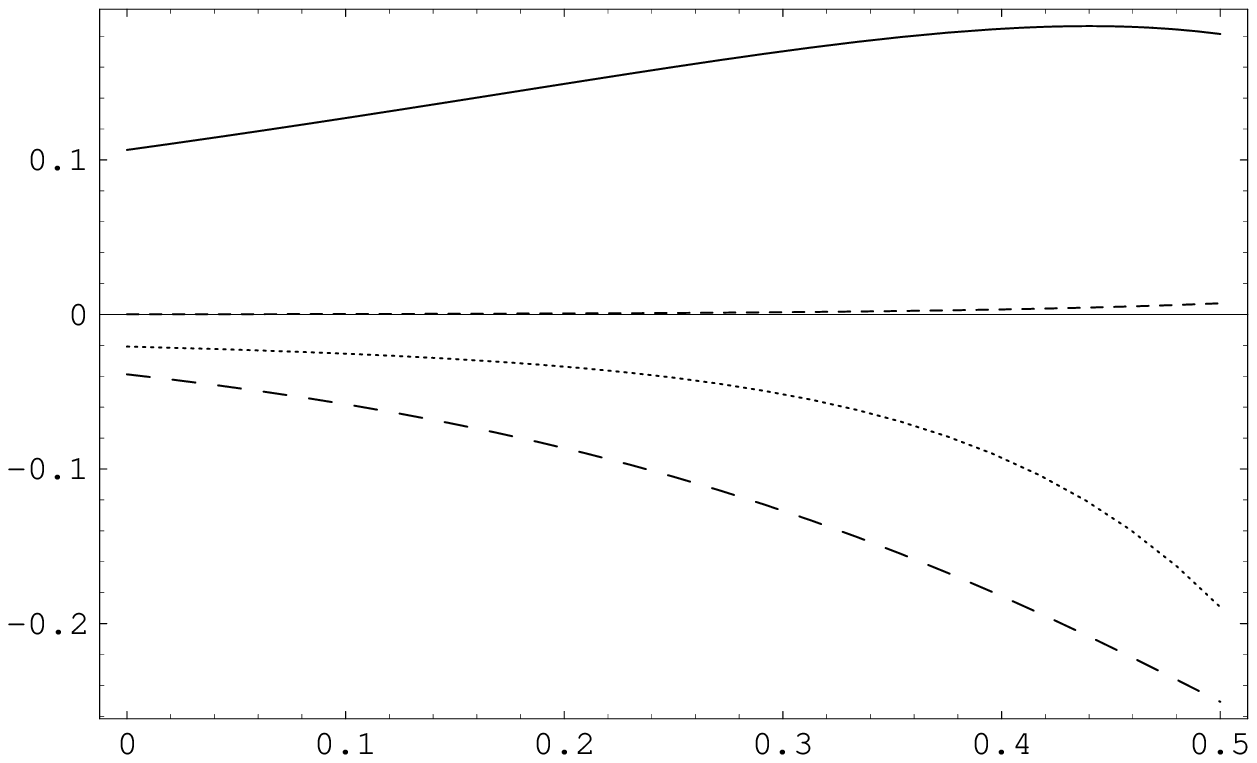}}
\\
{\hspace{0.5cm}$y$\hspace{8.5cm}$y$}
\\
{$ $}
\\
\includegraphics[width=3in]{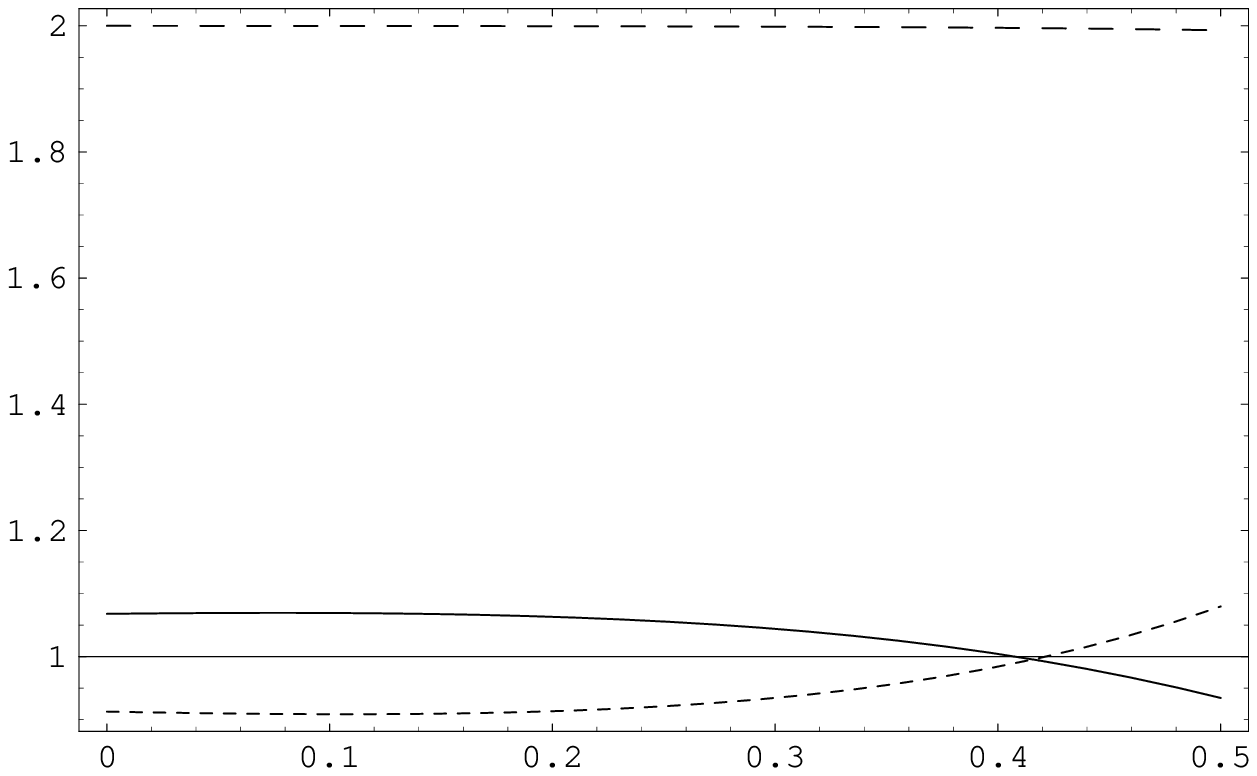}
\hspace{0,5cm}
\includegraphics[width=3.05in]{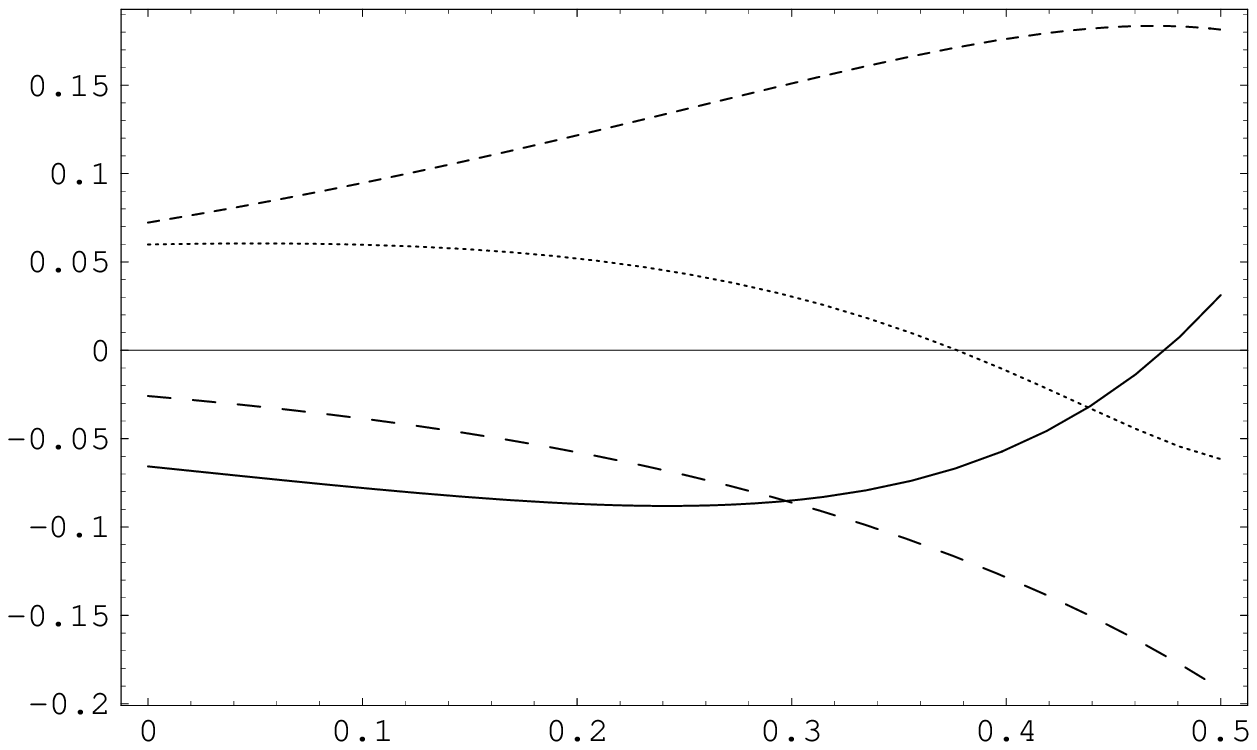}
\\
{\hspace{0.5cm}$y$\hspace{8.5cm}$y$}
\\
\caption{The graphs on the left contain the plot of 
the absolute values of the leading terms among $\eta_l$
(above) and $\xi_l$ (below) to $\mathcal{O}(\epsilon^2)$.
The graphs on the right show the subleading terms among
$\eta_l$ (above) and $\xi_l$ (below) and the
corresponding remainders $R_\eta$ and $R_\xi$ to 
$\mathcal{O}(\epsilon^2)$ (dotted lines).
All plots are for $\rho_i=2\cdot 10^{-1}\,M^4$
and $w_i=0.5$ and cover all the bulk between the two
branes.
\label{FappH2AC}}
\end{figure*}
\begin{subequations}
\be
\xi_1(y,\rho)&\equiv&
\frac{1}{n^2(y,t)}\,\frac{\ddot a(y,t)}{a(y,t)}
\label{F44terms1}
\\
\nonumber
\\
\xi_2(y,\rho)&\equiv&
-\frac{1}{b^2(y,t)}\,
\left(\frac{a'(y,t)}{a(y,t)}\right)^2
\label{F44terms2}
\\
\nonumber
\\
\xi_3(y,\rho)&\equiv&
-\frac{1}{b^2(y,t)}\,
\frac{a'(y,t)}{a(y,t)}\frac{n'(y,t)}{n(y,t)}
\label{F44terms3}
\\
\nonumber
\\
\xi_4(y,\rho)&\equiv&
\frac{1}{n^2(y,t)}\,\left(\frac{\dot a^2(y,t)}{a^2(y,t)}-
\frac{\dot a(y,t)\,\dot n(y,t)}{a(y,t)\,n(y,t)}\right)
\label{F44terms4}
\\
\nonumber
\\
\xi_5(y,\rho)&\equiv&
-\frac{k^2}{3}\,(\Lambda+U)
\label{F44terms5}
\\
\nonumber
\\
\xi_6(y,\rho)&\equiv&
-\frac{2}{3}\,k^2\,\omega^2\,b(y,t)\,
\left[b(y,t)-b_0\right]
\ ,
\label{F44terms6}
\ee
\end{subequations}
and 
\be
R_\xi(y,\rho)
\equiv \sum_l\xi_l
\ ,
\ee
with $R_\xi(y,\rho)=\mathcal{O}(\epsilon^3)$ for the 
same approximate solutions.
One may now assume that approximate metric functions computed
to $\mathcal{O}(\epsilon^2)$ are accurate approximations
of exact solutions to Eqs.~(\ref{Ein1}) if 
\be
|\eta_l|\gg |R_1|
\quad
{\rm and}
\quad
|\xi_l|\gg|R_2|
\ ,
\ee
for every term in Eqs.~(\ref{F00terms1})-(\ref{F44terms6})
evaluated to $\mathcal{O}(\epsilon^2)$.
(It should actually be sufficient to satisfy the above conditions
for the leading terms of each equation.)
Throughout this section, where unspecified, a natural
choice~\footnote{In terms of the fundamental scale $M$ one
has $m=M$, $k^2=M^{-3}$, $b_0=M^{-1}$, and $\omega^2=M^7$.} 
of dimensionful parameters is considered.
\par
Fig.~\ref{FappH2AC} shows the functions $\eta_l$ and $\xi_l$
evaluated to second order in $\epsilon$ and the corresponding
$R_\eta$ and $R_\xi$ for $\rho_i=2\cdot 10^{-1}\,M^4$ and
$w_i=0.5$. 
The box in the first row on the left shows the absolute
values of the leading terms $\eta_2$ and $\eta_3$, which
are roughly one order of magnitude larger than the remaining 
terms $\eta_l$'s displayed in the plot on the right along
with $R_\eta$ (the dotted line).
In the second row of Fig.~\ref{FappH2AC}, the modula of the
leading terms $\xi_2$, $\xi_3$ and $\xi_5$ are plotted on the
left while the remaining coefficients $\xi_l$ and $R_\xi(y)$
are presented on the right. 
Thanks to the relatively large amplitudes of the leading terms,
one can rely on the second order approximation even when the
$\rho_i$'s are not too small, implying that a great improvement
over first order results can be achieved in a regime where
$\rho_i\sim 10^{-2}$. 
In this case the errors produced by truncating the expansion
to $\mathcal{O}(\epsilon^2)$ are much smaller than one
percent.
\par
\begin{figure*}[ht]
\includegraphics[width=3in]{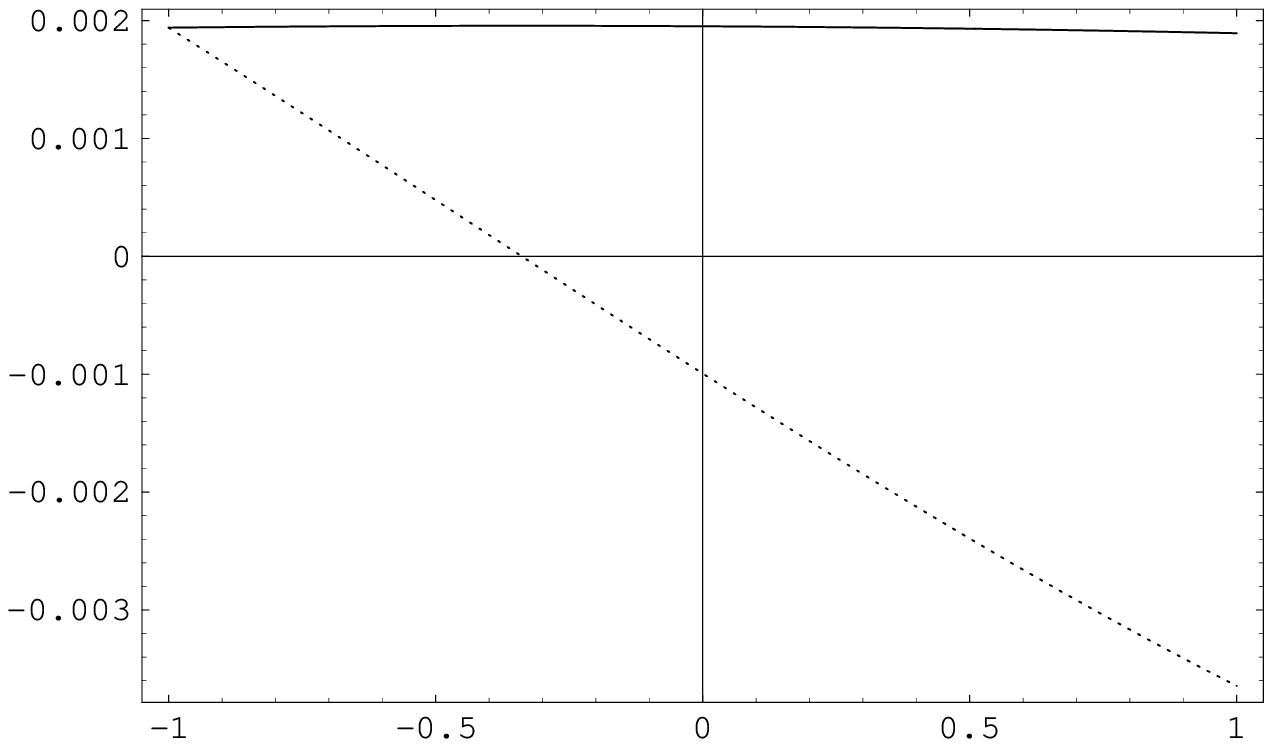}
\hspace{0,5 cm}
\includegraphics[width=3in]{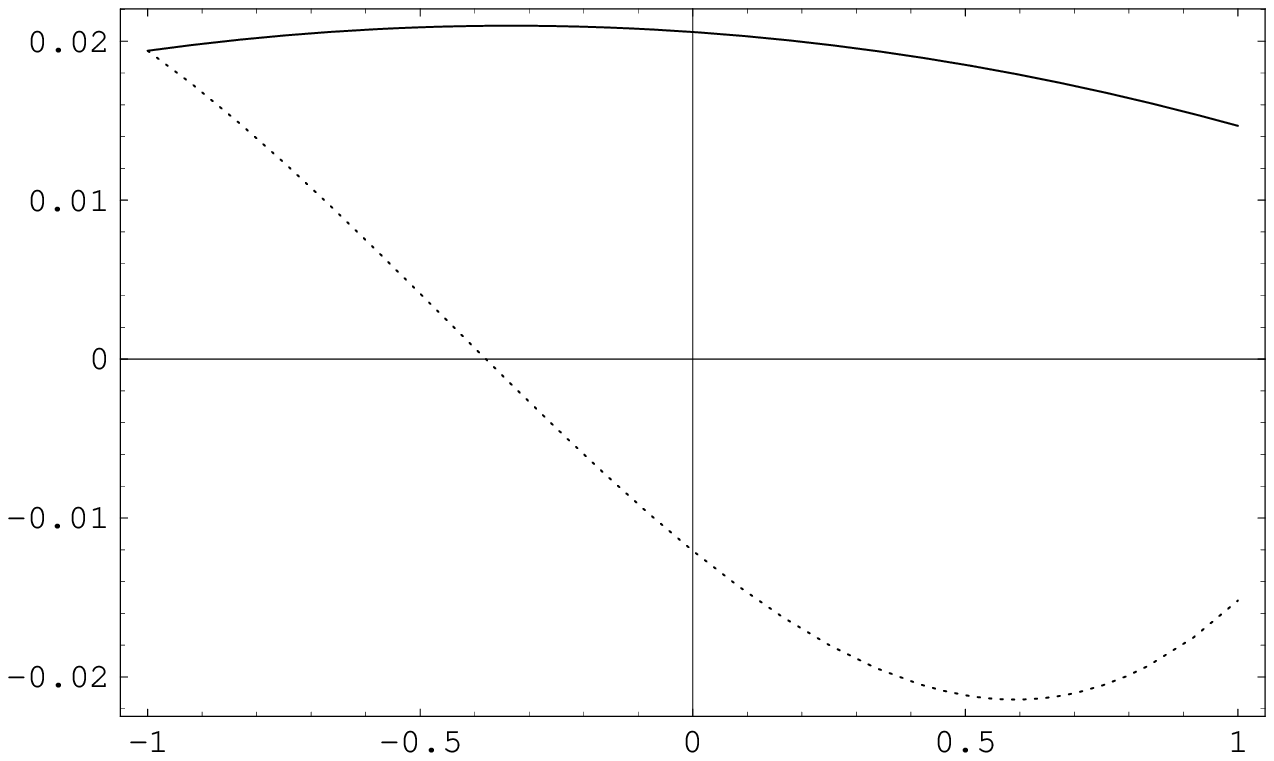}
\\
{\hspace{1cm}$w_n$\hspace{8cm}$w_n$}
\\
{$ $}
\\
\includegraphics[width=3in]{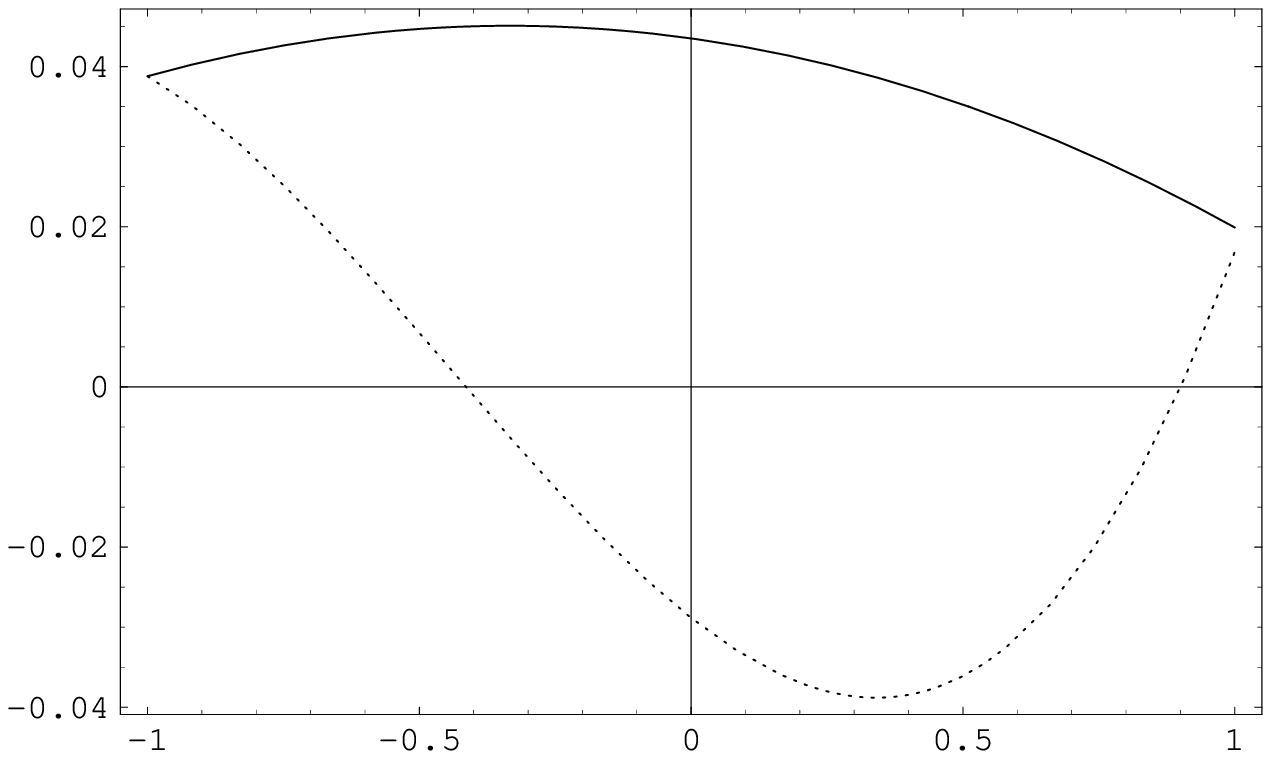}
\hspace{0,7 cm}
\includegraphics[width=2.9in]{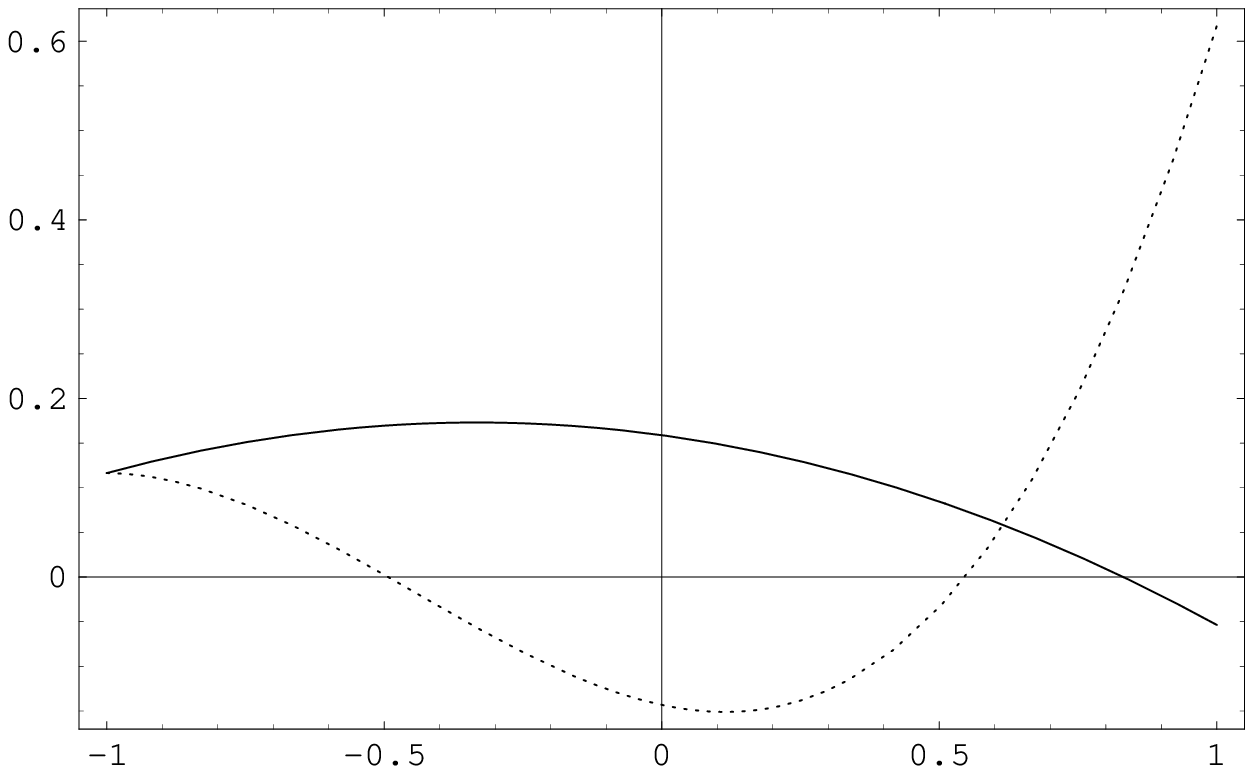}
\\
{\hspace{1cm}$w_n$\hspace{8cm}$w_n$}
\\
\caption{Plots of $H^2(1/2,\tau)$ (solid line) 
and $\ddot a(1/2,\tau)/a(1/2,\tau)$ (dotted line)
to $\mathcal{O}(\epsilon^2)$ at a given time,
as functions of $w_n$, for $\rho_n=10^{-2}\,M^4$,
$10^{-1}\,M^4$, $2\cdot 10^{-1}\,M^4$ and
$6\cdot 10^{-1}\,M^4$
(from top left to bottom right).
\label{Fvb1H2AC}}
\end{figure*}
Let us further consider the particular case $\rho_p=0$.
This choice is made in order to study the cosmological
consequences experienced on the negative tension brane
(at $y=1/2$) generated by second order terms proportional
to $\rho_n^2$.
We therefore use the proper time $\tau$ on the negative
tension brane~\footnote{Let us recall that this is
achieved by adopting the particular gauge choice
for $c_{n,i}^{(1)}$ and $c_{n,i}^{(2)}$ which sets $f_{n,i}^{(1)}(t)=f_{n,i}^{(2)}(t)=0$
and $c_{a,i}^{(1)}=c_{a,i}^{(2)}=0$.}.
In each graph in Fig.~\ref{Fvb1H2AC} we plot the squared
Hubble parameter on the negative tension brane of 
Eq.~(\ref{FrEqRhoPzero}) and the acceleration of
Eq.~(\ref{2ordnegAccEqRhoPzero}) for a given energy
density $\rho_n$ as a function of its equation of state
$w_n$. 
The plot on the left in the first row is for the small
density $\rho_n=10^{-2}\,M^4\ll M^4$ and shows a behavior
which is typical of standard four-dimensional cosmology.
In fact, only $w_n<-1/3$ leads to an accelerating phase. 
This trend is modified by higher densities, as it emerges
in the remaining graphs of Fig.~\ref{Fvb1H2AC}. 
In particular when $\rho_n=2\cdot 10^{-1}\,M^4$ 
(plot on the left in the second row) the second order
corrections seem to provide an accelerated regime
for $0<w_n<1$. 
The acceleration appears amplified when
$\rho_n=6\cdot 10^{-1}\,M^4<M^4$ (plot on the right
in the second row) or higher.
Note however that the Hubble parameter, a positive definite
quantity, constrains the region swept by $w_n$ which
is not allowed to reach unity. 
Finally the intermediate case is showed in the plot 
on the right the first row: due to the second order
effect, $H^2$ exhibits a dependence on $w_n$ otherwise
not present.
\par
\begin{figure*}[ht]
\includegraphics[width=3.1in]{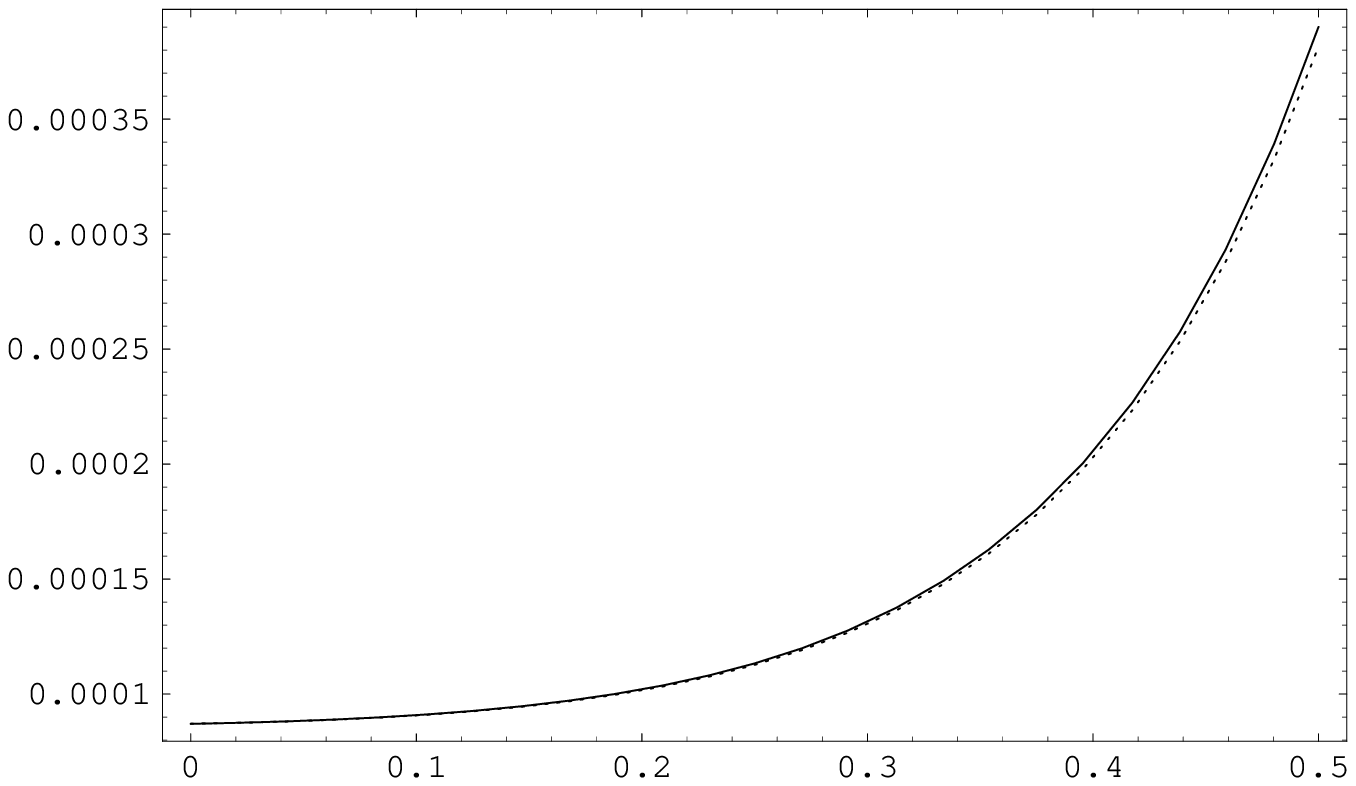}
\hspace{0,5 cm}
\includegraphics[width=3in]{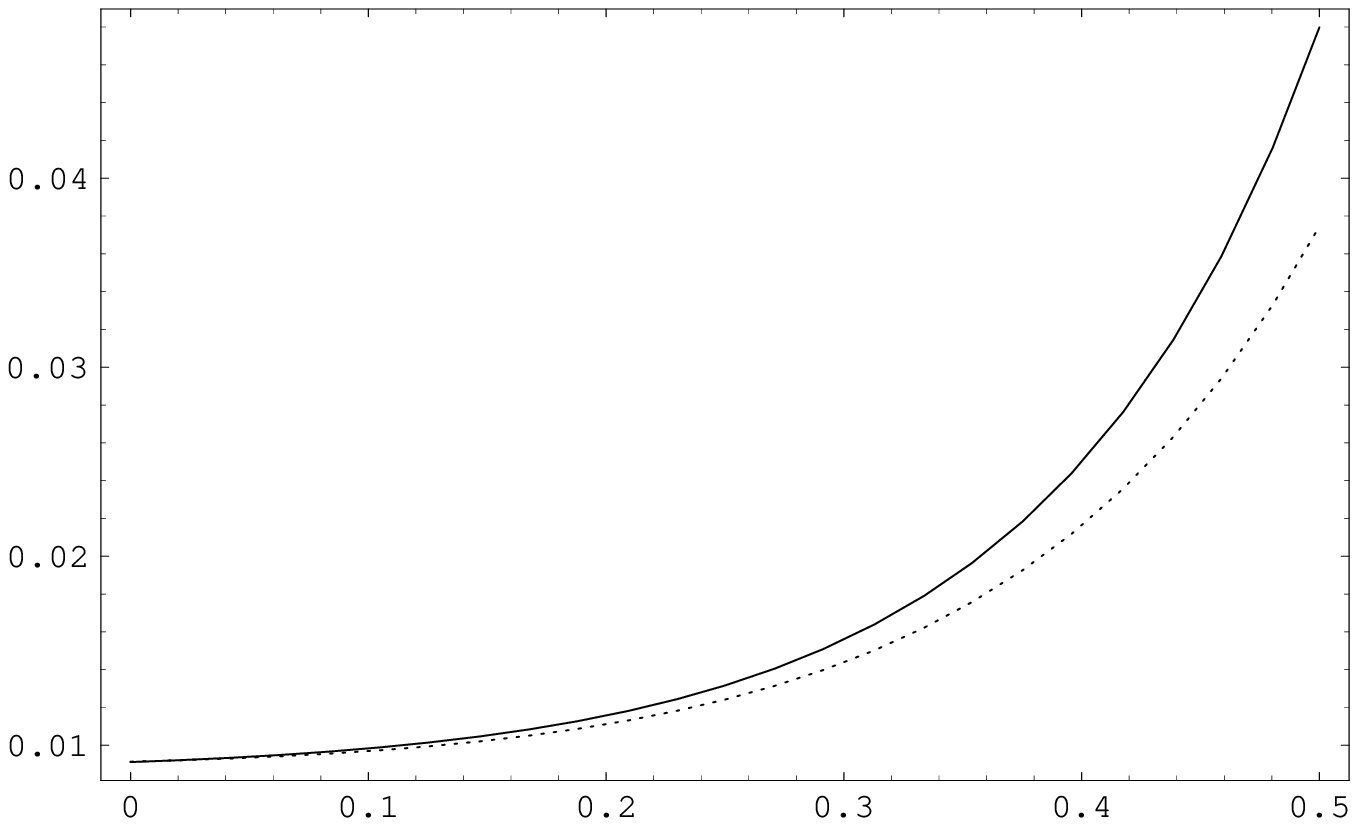}
\\
{\hspace{0.5cm}$y$\hspace{8.5cm}$y$}
\\
{$ $}
\\
\hspace{0.3 cm}
\includegraphics[width=2.92in]{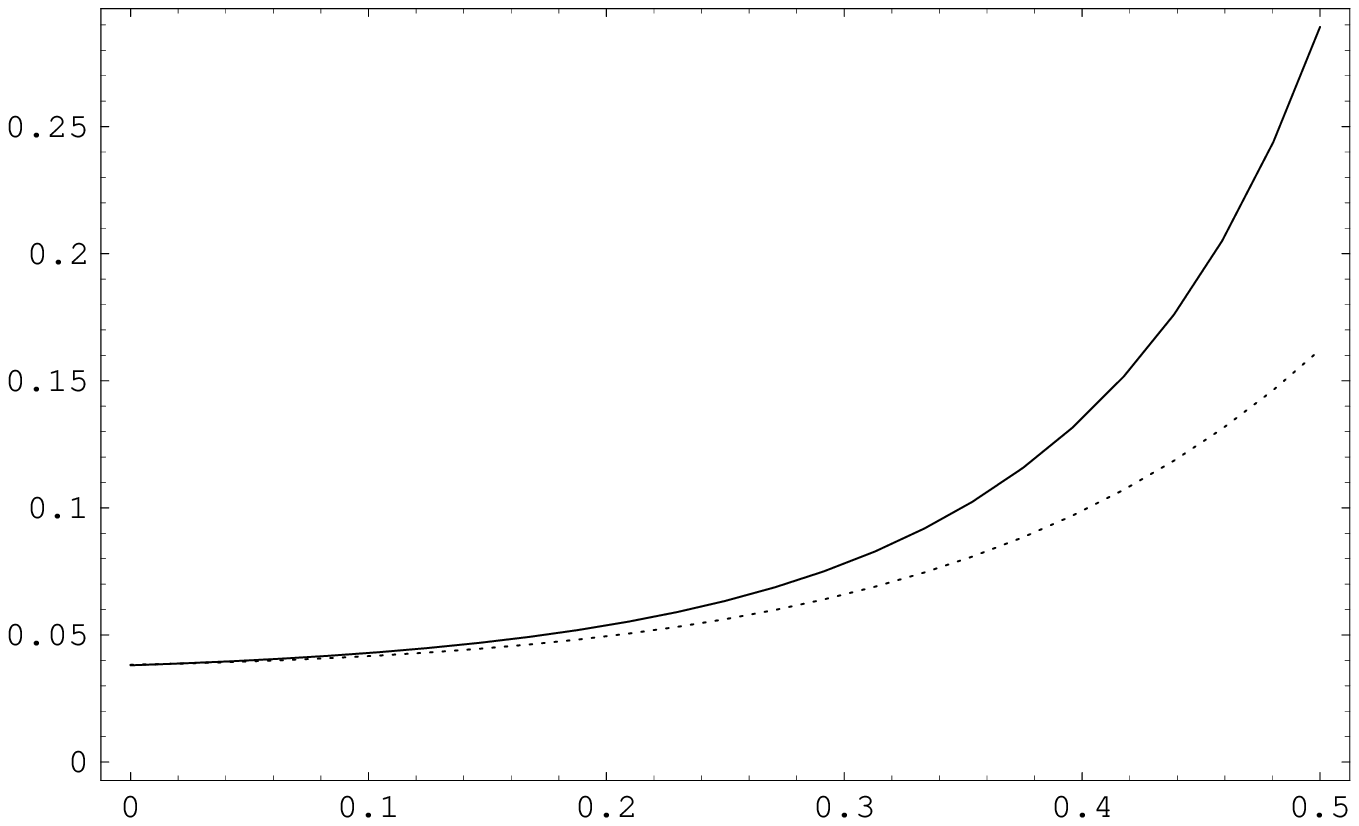}
\hspace{0,7 cm}
\includegraphics[width=2.92in]{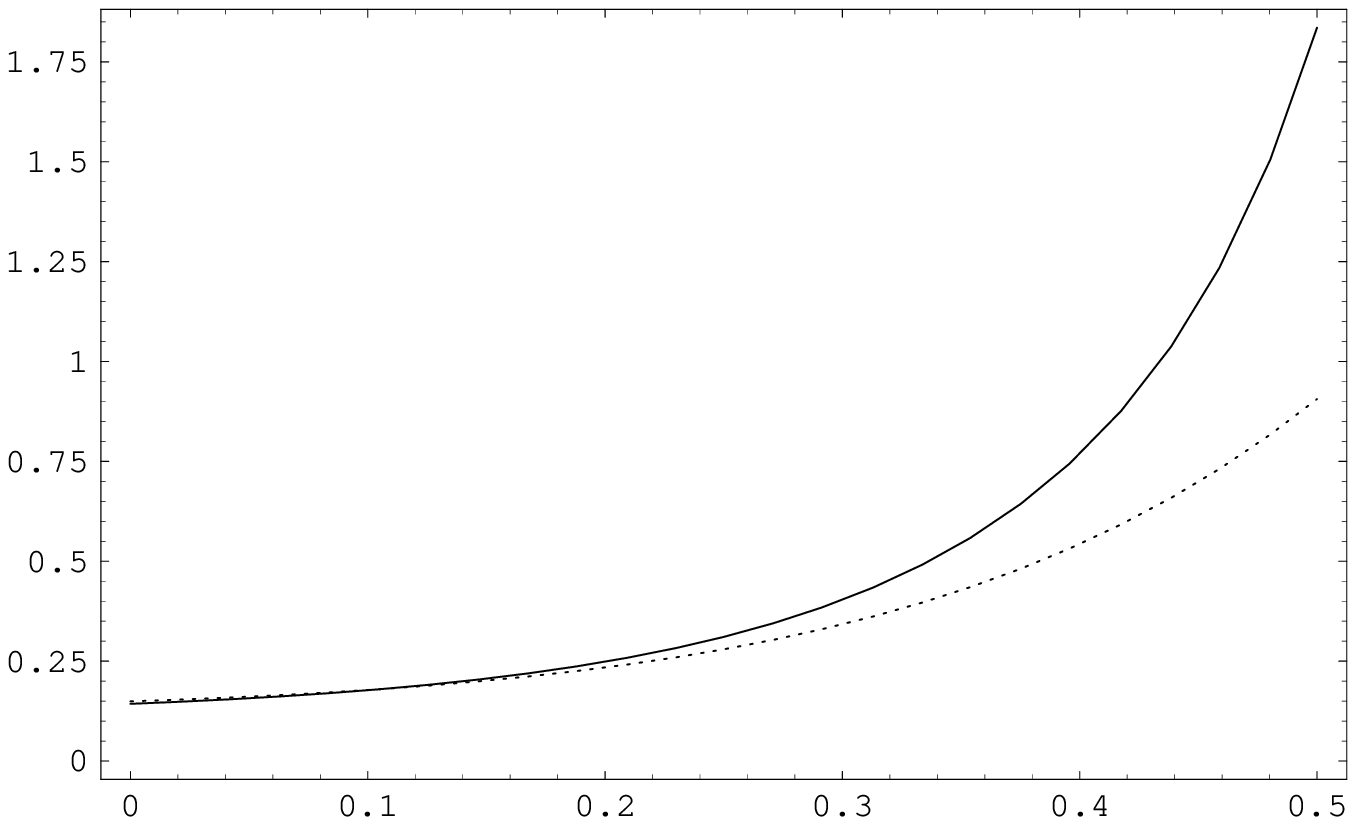}
\\
{\hspace{0.5cm}$y$\hspace{8.5cm}$y$}
\\
\caption{The two graphs in the first row show
the ratio between $R_\eta$ to $\mathcal{O}(\epsilon^2)$ and
$\eta_1\sim H^2$ to $\mathcal{O}(\epsilon)$ (dotted line)
and to $\mathcal{O}(\epsilon^2)$ (solid line) at a given time,
for $\rho_n=10^{-2}\,M^4$ and $w_n=0.95$ (left) and for
$\rho_n=10^{-1}\,M^4$ and $w_n=0.95$ (right). 
In the two graphs in the second row, the same ratios
are are given for $\rho_n=2\cdot 10^{-1}\,M^4$ and $w_n=0.95$
(left) and for $\rho_n=6\cdot 10^{-1}\,M^4$ and $w_n=0.65$ 
(right).
The plots cover all the bulk between the two branes.
\label{Fvb1ErrH2}}
\end{figure*}
\begin{figure*}[ht]
\includegraphics[width=3.1in]{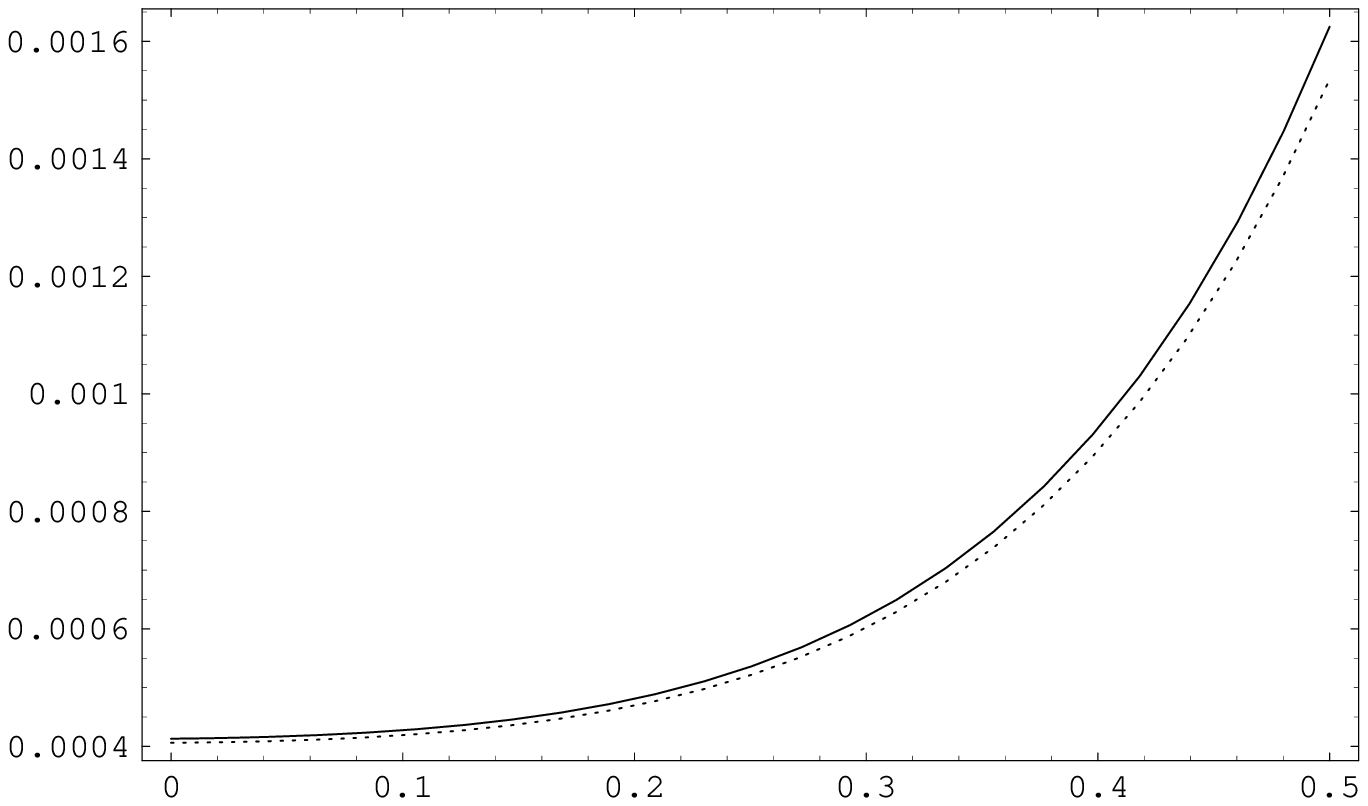}
\hspace{0,5 cm}
\includegraphics[width=3in]{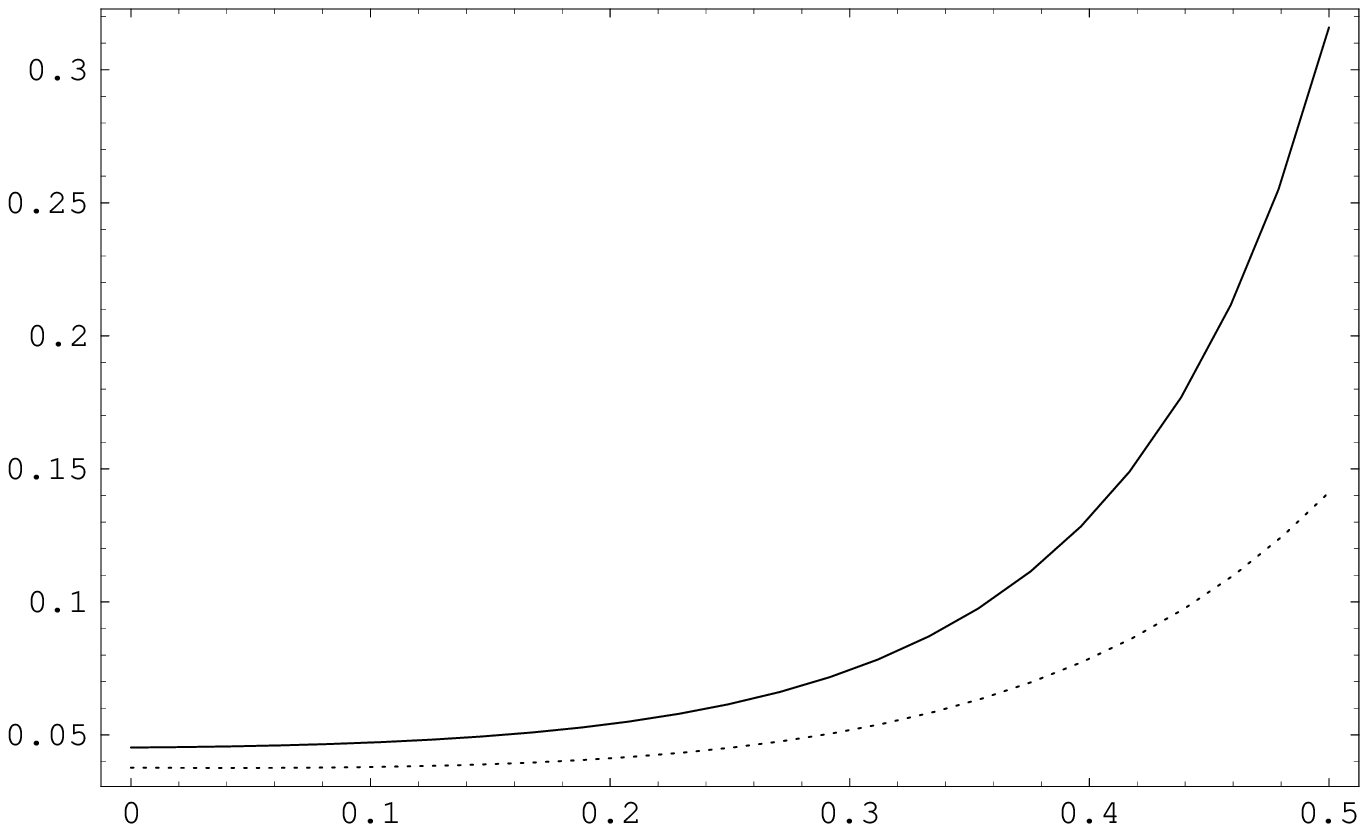}
\\
{\hspace{0.5cm}$y$\hspace{8.5cm}$y$}
\\
{$ $}
\\
\hspace{0.3cm}
\includegraphics[width=2.92in]{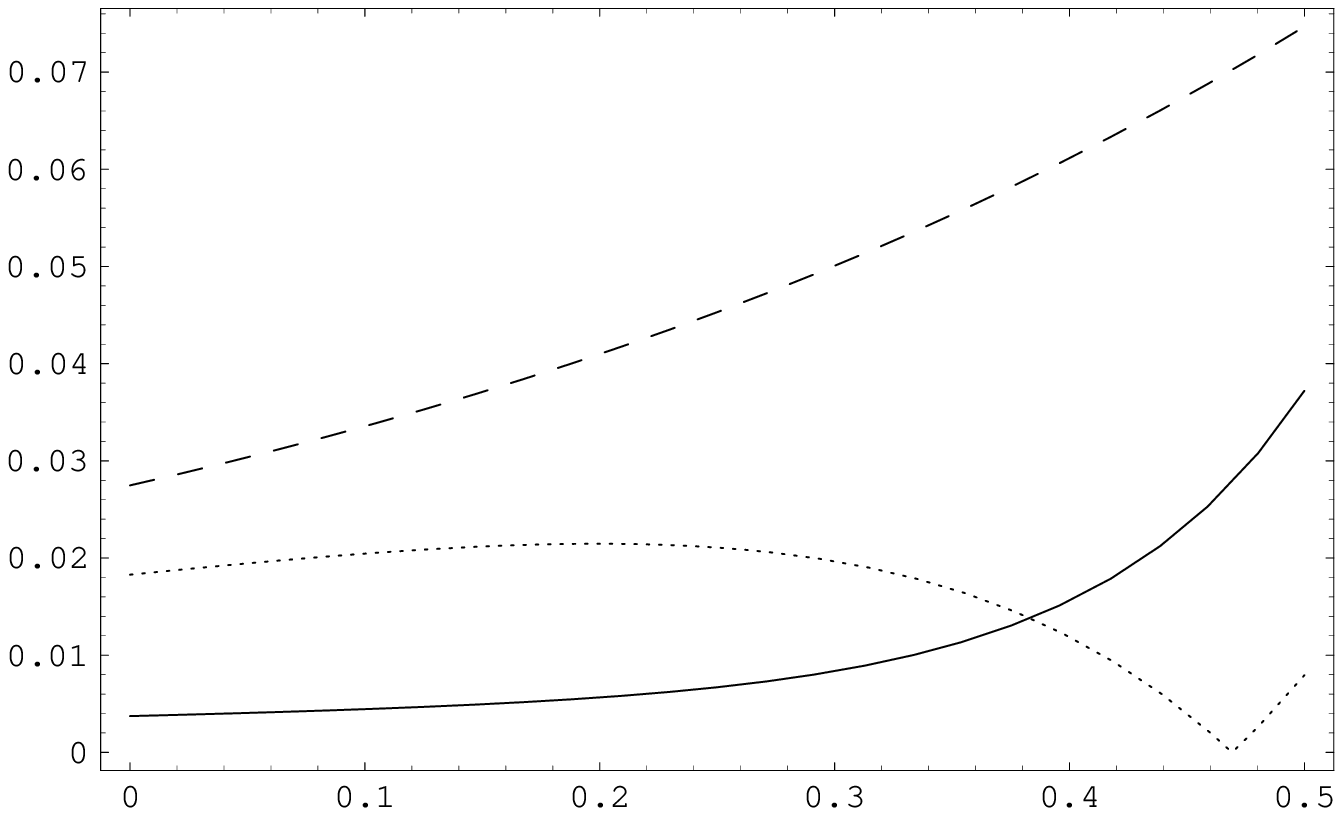}
\hspace{0,7 cm}
\includegraphics[width=2.92in]{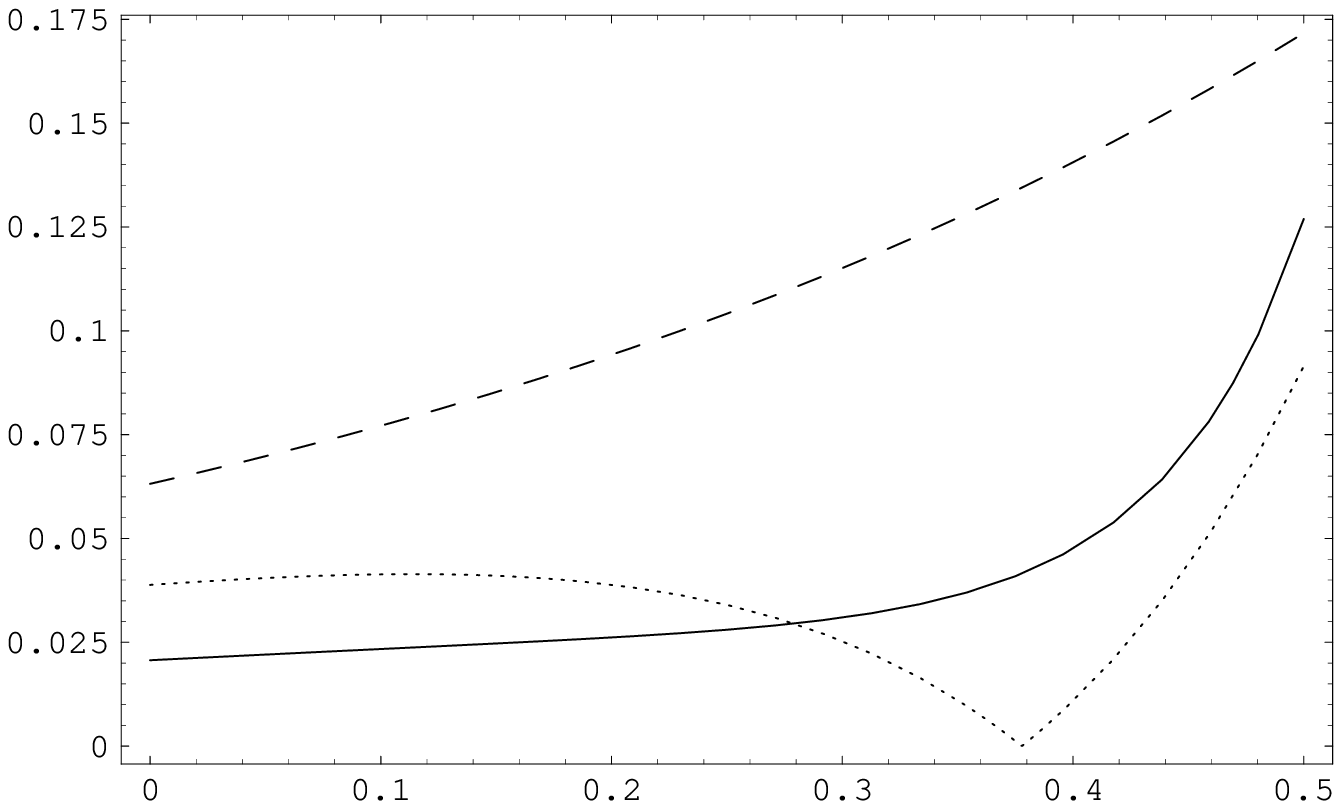}
\\
{\hspace{0.5cm}$y$\hspace{8.5cm}$y$}
\\
\caption{The two graphs in the first row show
the ratio between $R_\xi$ to $\mathcal{O}(\epsilon^2)$ and
$\xi_1\sim \ddot a/a$ to $\mathcal{O}(\epsilon)$ (dotted line)
and to $\mathcal{O}(\epsilon^2)$ (solid line) at a given time,
for $\rho_n=10^{-2}\,M^4$ and $w_n=0.95$
(left) and for $\rho_n=10^{-1}\,M^4$ and $w_n=0.95$
(right). 
In the two graphs in the second row, 
$R_\xi$ to $\mathcal{O}(\epsilon^2)$
(solid line) is compared to the modula of $\xi_1$
to $\mathcal{O}(\epsilon)$ (dashed line) and to
$\mathcal{O}(\epsilon^2)$ (dotted line) for
$\rho_n=2\cdot 10^{-1}\,M^4$ and $w_n=0.95$ (left),
and for $\rho_n=6\cdot 10^{-1}\,M^4$ and $w_n=0.65$ 
(right).
The plots cover all the bulk between the two branes.
\label{Fvb1ErrAC}}
\end{figure*}

The four regimes described above have to be tested
with particular accuracy because the acceleration
and Hubble parameter plotted in Fig.~\ref{Fvb1H2AC}
are not the leading terms in Eqs.~(\ref{Ein1}) and
could thus be comparable with the remainders.
In this case, it is somehow possible that the remainders
significantly modify the behavior.
We first note that $H^2\sim \eta_1$ and
$\ddot a/a\sim\xi_1$, as defined above in
Eqs.~(\ref{F00terms1}) and (\ref{F44terms1}), hence we
can use $\eta_1$ and $\xi_1$ in place of $H^2$ and
$\ddot a/a$ respectively.
We then plot in Fig.~\ref{Fvb1ErrH2} the ratio between
the remainder $R_\eta$ (evaluated to second order) and
the squared Hubble parameter $\eta_1$ evaluated to
first and second orders.
The four plots show this ratio for different choices
of $\rho_n$ and $w_n$ (the same as in Fig.~\ref{Fvb1H2AC} 
and in the same order) as a function of $y$.
In particular, we choose $w_n=0.95$ in order to explore
regions where the acceleration has an unconventional
behavior for $\rho_n=10^{-2}\,M^4$,
$10^{-1}\,M^4$, $2\cdot 10^{-1}\,M^4$ and $w_n=0.65$ for
$\rho_n=6\cdot10^{-1}\,M^4$. 
Apart from the last case, the corrections given by
the neglected terms cannot significantly modify $H^2$
on the negative tension brane.
In the first line of Fig.~\ref{Fvb1ErrAC},
the ratios between $R_\xi$ (to second order) and the
acceleration $\xi_1$ to first and second order
are analogously plotted.
For low enough energy densities, the second order
expressions appear to be good approximations.
On the other hand, in the second row it is shown
that, in the unconventional regime of the acceleration
(that is $\rho_n=2\cdot 10^{-1}\,M^4$ and $w_n=0.95$, and
$\rho_n=6\cdot 10^{-1}\,M^4$ and $w_n=0.65$) the two terms
are of the same order of magnitude.
This shows that one should go beyond $\mathcal{O}(\epsilon^2)$
in order to determine the true behavior of $\ddot a/a$
for such equations of state. 
\section{Conclusions}
\label{VI}
We have computed approximate cosmological solutions of
five-dimensional Einstein equations for Randall-Sundrum
models in the presence of a radion effective potential.
The calculations were performed up to the second order in
the energy densities of the matter on the branes and assuming
an adiabatic evolution of the system.
Our approach differs from Ref.~\cite{vinet} in that we do
not include a specific bulk field to achieve stabilization, and
is therefore more general.
Interestingly, their results are recovered
in the limit of very large warp factor and radion mass.
For the RS~I model with matter localized only on the negative
tension brane, we found negligible corrections for the Hubble
parameter in the case of radiation or cosmological constant,
thus supporting one of the main results of Ref.~\cite{vinet}.
For RS~II, we found negligible corrections for the equations
of state just described and in the limit when the distance
between the branes is taken to infinity.
\par
On inspecting our results, we finally found some evidence of an
accelerating phase for a wider range of values of the
equation of state $p_n= w_n\,\rho_n $ on the negative
tension brane if the distance between the branes is finite.
However, one should then carry the computation to higher
orders, since such an effect appears near the limit of
validity of our perturbative expansion.

\end{document}